\documentclass[traditabstract]{aa}                                    
\usepackage{graphicx,hyperref}
\usepackage{txfonts}
\usepackage{natbib}
\begin{document}
   \title{Roche-lobe filling factor of mass-transferring red giants -- the PIONIER view}
\titlerunning{PIONIER view of mass-transferring red giants}
\authorrunning{H.M.J. Boffin et al.}

   \author{Henri M.J. Boffin\inst{1}
    \and
           M. Hillen\inst{2}
           \and
           J.P. Berger\inst{3}
                     \and
           A. Jorissen\inst{4}
           \and
           N. Blind\inst{5}
            \and\\
           J.B. Le Bouquin\inst{6}  
           \and   
                    J. Miko{\l}ajewska\inst{7}
          \and 
          B. Lazareff\inst{6}
         }

   \institute{ESO, Alonso de C\'ordova 3107, Casilla 19001, Santiago, Chile\\
              \email{hboffin@eso.org}
      \and
            Instituut voor Sterrenkunde, KU Leuven, Celestijnenlaan 200D, B-3001 Leuven, Belgium
            \and
            ESO, Karl-Schwarzschild-Strasse 2, D-85748 Garching bei M\"unchen, Germany   
             \and
            Institut d'Astronomie et d'Astrophysique, Universit\'e Libre de Bruxelles, Campus Plaine C.P. 226, Bd du Triomphe, B-1050 Bruxelles, Belgium
            \and
            Max Planck Institute for Extraterrestrial Physics, Giessenbachstrasse, D-85741  Garching, Germany
            \and
            UJF-Grenoble 1/CNRS-INSU, Institut de Plan\'etologie et d'Astrophysique de Grenoble (IPAG) UMR 5274, Grenoble, France 
            \and
            Copernicus Astronomical Center, Bartycka 18, PL 00-716 Warsaw, Poland
                        }

   \date{Received 4 December 2013; accepted 31 January 2013}

 
 \abstract
{Using the PIONIER visitor instrument that combines the light of the four Auxiliary Telescopes of ESO's Very Large Telescope Interferometer, we measure precisely the diameters of several symbiotic and related stars: HD~352, HD~190658, V1261~Ori, ER~Del, FG~Ser, and AG~Peg. These diameters -- in the range of 0.6 to 2.3 milli-arcseconds -- are used to assess the filling factor of the Roche lobe of the mass-losing giants and provide indications on the nature of the ongoing mass transfer. We also provide the first spectroscopic orbit of ER~Del, based on CORAVEL and HERMES/Mercator observations. The system is found to have an eccentric orbit with a period of 5.7 years. In the case of the symbiotic star FG Ser, we find that the diameter is changing by 13\% over the course of 41 days, while the observations of HD~352 are indicative of an elongation. Both these stars are found to have a Roche filling factor close to 1, as  is most likely the case for HD 190658 as well, while the three other stars have factors below 0.5--0.6. Our observations reveal the power of interferometry for the study of interacting binary stars -- the main limitation in our conclusions being the poorly known distances of the objects.}


   \maketitle

\section{Introduction}

Symbiotic stars show in their spectra the blended characteristics of a cool star (generally a K or M giant), a hot star (a white dwarf in most cases), 
as well as emission lines coming from a high-excitation nebula. With orbital periods in the range of a few hundred to a thousand days, they are thought to be among the interacting binary stars with the longest periods, in which the mass-losing red giant is transferring mass to its hot companion \citep{mikolajewska_2007}. 
Their study has important implications for a wide range of objects, such as Type Ia supernovae, barium stars, the shaping of planetary nebulae, and compact binaries such as cataclysmic variables \citep{podsiakolski_2007}. 

A critical question related to symbiotic stars is whether mass transfer is taking place through a stellar wind or Roche lobe overflow \citep{mikolajewska_2012}, i.e. what is the Roche-lobe-filling factor of the giant in those systems. There is a well-known apparent contradiction between the radius derived from the rotational velocities (when assuming synchronisation), which in most cases indicate that the giant fills about only half its Roche lobe, and the fact that many of the symbiotic stars show the clear signature of ellipsoidal variations in their light curve, indicative of a much larger filling factor. 
This paper aims at addressing this question in the most direct way.

\citet{blind11} have shown the power of interferometry when studying the symbiotic star SS Lep. Using PIONIER/VLTI data, they fully constrained the orbit of the system and determined the mass of each component. They also revised the M giant's size to lower values than previously found, proving that the system was not currently undergoing a {\it strict}\footnote{There is indeed still the possibility to have wind Roche lobe overflow, or that the Roche lobe radius is reduced with respect to the canonical value, in case there is a strong mass loss \citep[see ][for more details]{blind11}.} Roche lobe overflow, the giant's filling factor being about 86\%. Significant mass transfer did happen in this system, however,  the mass of the most evolved star  being less than half the mass of its companion. 

It is not possible to always obtain as many data as in the case of SS Lep (e.g., in most symbiotic systems, the hot star would not have a signature in the infrared) and thus to constrain as much the system. However, there are several systems for which it is possible to determine, with great accuracy, the diameter of the mass-losing giant. This, combined with previous data, could already answer the critical question raised above and more particularly, constrain the Roche lobe filling factor. Using the PIONIER/VLTI  interferometric instrument, we have measured therefore the diameter of several symbiotic and related stars. The observations are presented in Sect.~\ref{sec:obs}, and we discuss in turn each system in Sect.~\ref{sec:sys}.


\section{Observations, data reduction and results}
\label{sec:obs}

We observed 6 symbiotic or related stars with the four 1.8-m Auxiliary Telescopes of ESO's Very Large Telescope Interferometer, using the PIONIER visitor instrument \citep{berger_2010,2011A&A...535A..67L} in the $H$-band on the nights of 3 March, 3 July and 13 August 2012 (see Table~\ref{tab:diam}).   
For all targets, except V1262~Ori, we used the prism in low resolution (SMALL) which provides a spectral resolution $R\sim15$, the fringes being sampled over three spectral channels. For  one of our  brightest objects in $H$, V1262~Ori, we used the ``high resolution'' (LARGE) mode, which provides a spectral resolution $R\sim40$, the fringes being sampled over seven spectral channels. 
The large VLTI configuration A1-G1-I1-K0 was used, leading to baselines of 47 (H1-I1 and K0-I1), 
80 (A1-G1), 91 (G1-K0), 107 (A1-I1) and 129 metres (A1-K0). We provide in the online Table~\ref{tab:calibPIONIER} the list of calibrators used.

\begin{figure}
   \centering
\includegraphics[width=9cm]{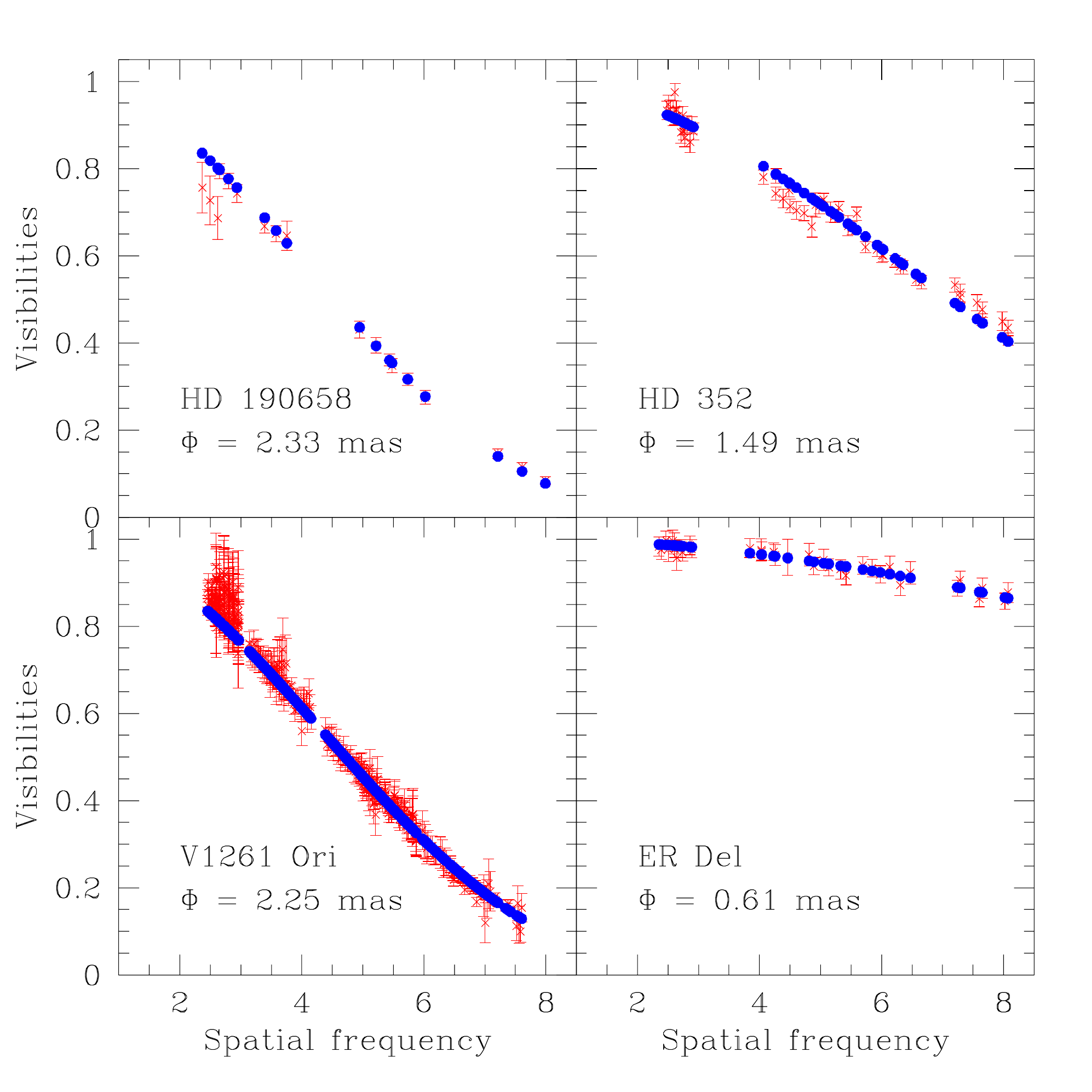}
      \caption{PIONIER squared visibilities for the stars HD~190658, HD~352, V1261~Ori, and ER Del as a function of the spatial frequency in 1/rad ($\times 10^7$). The data points are shown in red with error bars, while the uniform disc model that provides the best fit is indicated with blue solid dots.
              }
         \label{fig:diam1}
   \end{figure}

\begin{figure}
   \centering
\includegraphics[width=9cm]{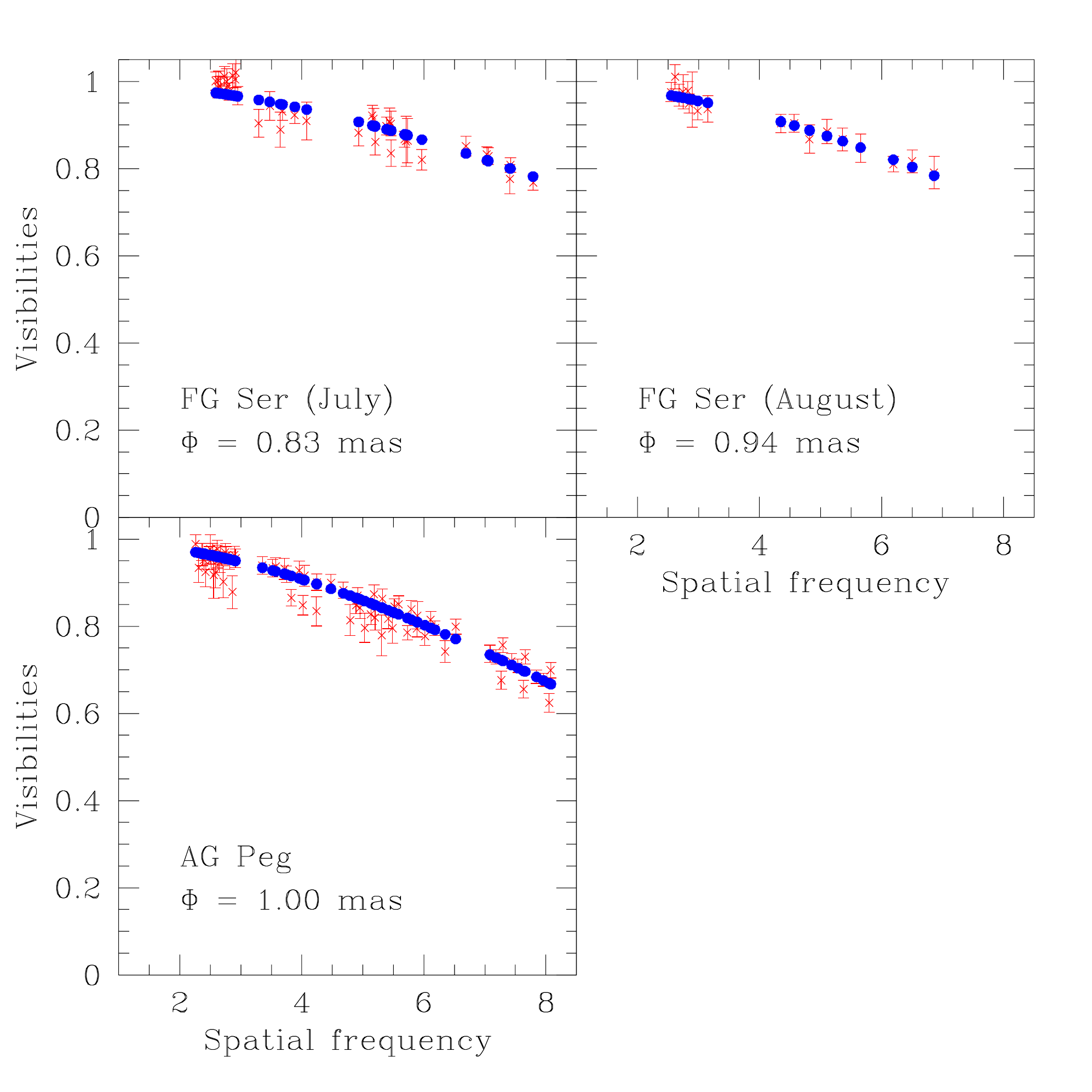}
      \caption{Same as Fig.~1 for the stars FG Ser (on 3 July and on 13 August 2012) and AG Peg.
              }
         \label{fig:diam2}
   \end{figure}

Data reduction was done in the usual way with the {\tt pndrs} package presented by Le Bouquin et al. (2011) and the {\tt LITpro} software\footnote{{\tt LITpro} is available from \url{http://www.jmmc.fr/litpro\_page.htm}} \citep{Litpro} was used to fit a uniform disc to the visibilities and closure phases. Our results are shown in Figs.~\ref{fig:diam1} and ~\ref{fig:diam2}, as well as in Table~\ref{tab:diam}.  As can be seen in this table, where the last column indicates the resulting reduced $\chi^2$, for all stars but HD 352, the assumption of a uniform disc leads to a very good fit.  The stars that are studied in this paper are all known to be single-lined spectroscopic binaries and it is thus safe to assume that there should be no contribution from the companion to the total flux. This is even more so in the $H$-band, as the companion is always a hot star. Our observations should thus mostly reveal only the primary giant, hence, our attempt to only fit a uniform disc to represent it. This conclusion is further strengthened by the fact that we did not detect any significant non-zero closure phases. There is a possible caveat, though, as revealed by the observations of SS Lep \citep{blind11}. The fact that the mass loss in symbiotic stars is much higher than in normal giants, leading to the observed activity, may result in either detectable circumstellar matter or a circumbinary disc. We have thus to make sure that there is no additional background present in our data.  
We therefore show in online Figs.~\ref{Fig:chi2map190} to \ref{Fig:chi2mapagpeg} the $\chi^2$ maps corresponding to models where a background is added to the uniform disc. These maps clearly show that the assumption of no background leads to the minimum $\chi^2$, i.e. the best fit. 
 In addition, we have verified that for all our targets but HD~352 (see below), using an elongated disc or Gaussian does not provide a better fit to our data.
The errors, $\sigma$, quoted in Table~\ref{tab:diam} have been conservatively calculated, based on the following:
$$\sigma^2 =  N_{\rm sp}\, \sigma_{\rm litpro}^2   +  0.0001\,  \Phi^2 \,$$
where $\Phi$ is the diameter in milli-arcseconds (mas). The first term takes into account the fact that the $N_{\rm sp}$ spectral channels of PIONIER are almost perfectly correlated, while LITpro assumes that points are independent, and the second term (a relative error of 1\%) arises from the fact that the wavelength calibration is precise at the 1\% level only, which leads to a similar uncertainty on the diameter. As mentioned above, for all our targets, $N_{\rm sp}$=3, except for V1262~Ori, where $N_{\rm sp}$=7.


\begin{table*}[htbp]
   \centering
      \caption{Measured diameter of our target stars.}
   \begin{tabular}{llllllccclc} 
\hline\hline
      \multicolumn{2}{c}{Star designation}  & ~~$V$ &~~$J$ & ~~$H$ & ~~$K$ & Date 	& Diameter 	& Error 	&$\chi^2_{\rm red}$\\
      		& 				& 		& &&	& & (mas) 		& (mas) 	& \\
      \hline
V1472~Aql & HD 190658 &  6.406 &  2.825 & 1.948 & 1.674 & 2012-07-03 & 2.33 & 0.03 & 0.82\\
AP Psc &HD 352 & 6.22 & 3.844 & 3.015 &  2.877 & 2012-08-13 & 1.49 & 0.02 & 2.48\\
& & & & & & &  1.38 $\times$ 1.60 & 0.02 &  1.25\\
      V1261~Ori  & HD 35155 & 6.87 &  3.336 & 2.415 & 2.138 & 2012-03-03  & 2.25 & 0.08 &0.87  \\
ER Del & -- & 10.39 &  6.185 & 5.337  & 4.987  & 2012-08-13 & 0.61 & 0.04 &0.80 \\
      FG Ser 	& AS 296 & 11.7 & 5.907 & 4.865 & 4.395 & 2012-07-03 	& 0.83	&	0.03	&0.69	 \\
 	&  & &  & & & 2012-08-13 	& 0.94 	&	0.05  	&0.26	 \\
	      AG Peg & HD 207757 & 8.65 &  5.001 & 4.371 & 3.851 & 2012-08-13 & 1.00 & 0.04 &1.31\\
      \hline
   \end{tabular}
  \label{tab:diam}
\end{table*}

\section{Discussion}
\label{sec:sys}

In this section, we now look at each of our target stars in turn, starting from those for which most data are available.
In our discussion, except for FG Ser, we will neglect any effect of extinction. We computed for all our targets the visual extinction, $A_V$, given by version 2.0.5 of the EXTINCT subroutine \citep{Hakkila97}, and  the resulting extinction in the $K$-band, using the usual conversion $A_K=0.114~A_V$. For all our targets, except FG Ser and HD~190658, we found $A_K$ to be of the order of 0.01 mag only. For HD~190658, we get $A_K=0.06$, which can also be neglected, while for FG Ser, we derive $A_K=0.185$~mag and we will take this into account in our discussion. 

Fitting our visibilities, we obtain uniform disc diameters. In principle, we should convert these to limb-darkened diameters, as these are generally used in computing luminosities.
However, the conversion factor from a uniform disc to a limb-darkened disc is in the range 1.02--1.04 \citep[see, e.g.,][and references therein]{blind11}.
The effect of neglecting extinction and of using a uniform disc instead of a limb-darkened disc are much smaller than the errors resulting from the distance, which are around 15\% or (much) larger,
and there is no need to introduce additional complications and unknowns.

\subsection{HD~190658}
HD~190658 (V1472~Aql, HR~7680, HIP~98954) is a M2.5 III star in a binary system with a rather short orbital period ($P=198.716$ days; \citealt{1982A&A...105..318L}) -- the second smallest in the sample of \citet{Famaey2009}. It is in a sense a sister system to the symbiotic star SS Lep, with a similar period, and therefore deserves further study.  It should be noted, though, that unlike SS Lep, this system is not an Algol, nor a double-lined spectroscopic binary and does not show any circumstellar or circumbinary material. In this sense, it is much  easier to model. 

{\it Hipparcos} measurements show photometric variations with about half the orbital period ($P= 100.37$~d) and 0.16 mag amplitude, which led \citet{1997IBVS.4501....1S} to conclude that the object is an eclipsing or ellipsoidal variable star.  
The fact that this period is almost exactly half the orbital period further strengthens this.

We can use the BC$_K$ vs. $(J-K)$ relation of \citet{2010A&A...524A..87K} to derive from our measured angular diameter the distance-independent effective temperature of the star, $T_{\rm eff}=3263\pm35$~K. 
Coupling the parallax $\varpi = 7.92 \pm 1.07$ mas \citep{2007A&A...474..653V} with our measured diameter of 2.33 mas, we derive a linear radius of 31.6$\pm$4.3 R$_\odot$.  

As the system presents detectable ellipsoidal variations, its inclination can be assumed to be rather high, typically between  50 and 90 degrees. If we take the latter value and 
assume\footnote{One needs a mass above 0.8-0.9~M$_\odot$ to let it become a giant in a Hubble time. On the other hand, given that its heliocentric velocity of $-110$ km~s$^{-1}$ hints at an old population, the giant cannot be too massive either.} that the giant has a mass between 0.8 and 3~M$_\odot$, we can then estimate companion masses between 0.40 and 0.88~M$_\odot$, and Roche lobe radii between 73 and 110~R$_\odot$, respectively, using the measured spectroscopic mass function $f(m) = 0.0449$~M$_\odot$ derived by \citet{1982A&A...105..318L}. Thus the filling factor of the giant is thought to be between 0.29 and 0.43. Using $i=50^{\circ}$ instead of $90^{\circ}$ would increase the filling factor by 3\% only. 

Thus, strangely enough, if we were to use the parallax from \citet{2007A&A...474..653V}, we would have to conclude that this star is far from filling its Roche lobe and the origin of the ellipsoidal variations would be quite puzzling. The absolute magnitude we deduce from using this parallax (M$_{\rm bol}=-0.25$ or $L=100$~L$_\odot$) also seems very low when compared to evolutionary tracks, and would imply the star to have a mass well below 1~M$_\odot$, which as mentioned above is unlikely (see Fig.~\ref{fig:hrdiag}).

D. Pourbaix (priv.~comm.) has reprocessed the original {\it Hipparcos} Intermediate Astrometric Data 
\citep[IAD; ][]{vanLeeuwen1998} and found that taking into account the effect of orbital motion would lead to a parallax of 2.6$\pm$0.9 mas, while if one assumes the photocentre to vary with half the orbital period (i.e. the ellipsoidal variation one) one gets a parallax of 
2.12 mas. We are thus led to conservatively conclude that the true parallax is given by $\varpi = 2.4 \pm 1.0$ mas. The star is thus three times farther away from us than previously thought! In this case, we derive a linear radius of  $104 \pm 56$~R$_\odot$! This means that the star is filling between 43 and 100\% of its Roche lobe. Using this parallax, we find a bolometric magnitude, $M_{\rm bol}=-2.6\pm1.0$, or a luminosity, $L=1100^{+1100}_{-700}$~L$_\odot$, much more in line with stellar evolution models.

\begin{figure}
   \centering
\includegraphics[width=9cm]{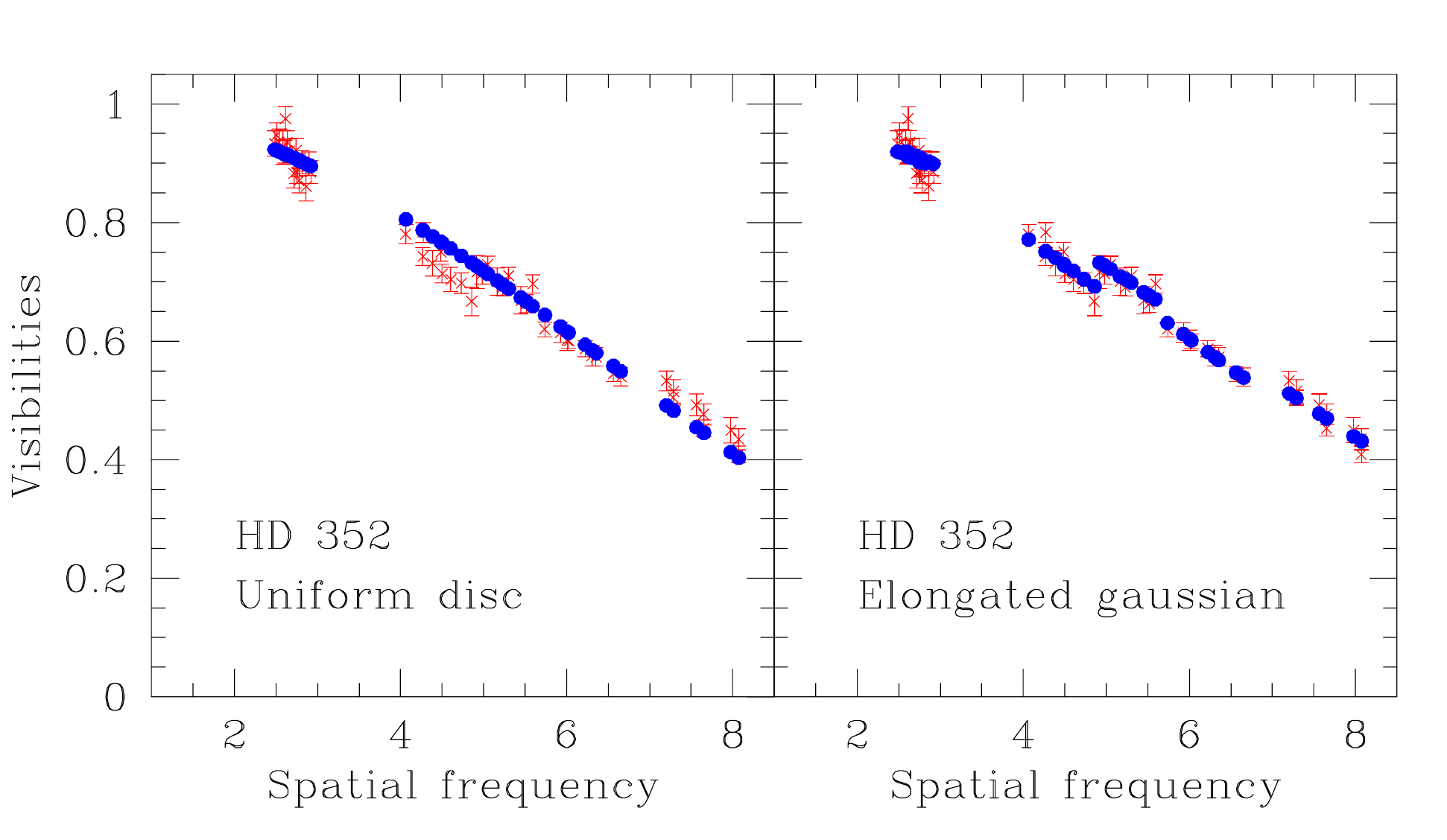}
      \caption{Squared visibilities for HD 352 as observed with PIONIER (red points with error bars), compared to a model with a uniform disc (blue dots; left panel) and an elongated Gaussian (right panel). The spatial frequency is given in 1/rad ($\times 10^7$). 
              }
         \label{fig:hr352}
   \end{figure}

\begin{figure}
   \centering
\includegraphics[width=9cm]{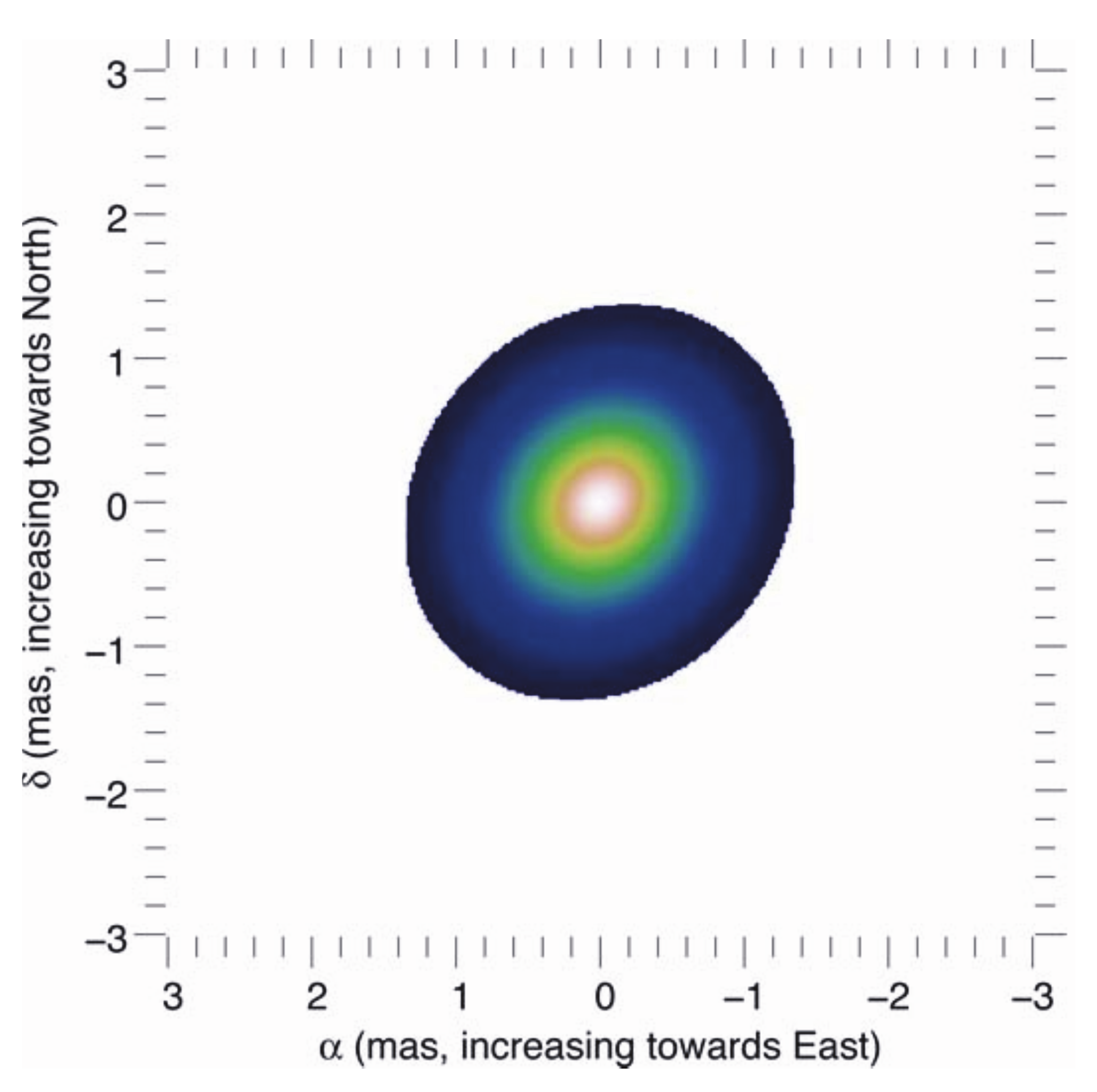}
      \caption{Model image of the elongated Gaussian that best fits the PIONIER visibilities for HD 352: the elongation ratio is $1.156 \pm 0.026$, the star is 1.38$\times$1.6 mas wide, with the major axis position angle being at $138 \pm 4$ degrees. 
                }
         \label{fig:hr352img}
   \end{figure}
   
\subsection{HD~352}
HD~352 (HR~14, AP Psc, 5 Cet, HIP~664) is, according to {\it SIMBAD}, an eclipsing binary of $\beta$ Lyr type, i.e. a semi-detached system. Its spectral type is indicated as K2--4 III.

\citet{1984IBVS.2589....1L} found the star to present ellipsoidal variations with amplitude $\sim 0.2$ mag, which was also found by {\it Hipparcos}. \citet{1985PASP...97..355B} provided a revised orbit with an orbital period of 94.439 days and \citet{1986IBVS.2952....1E} reported the rotational velocity to be $v \sin i=22\pm3$ km s$^{-1}$. The same authors later claimed \citep{1988AcA....38..353E} that the main-sequence companion is accreting mass from a Roche-lobe overflowing giant at a rate of $5 \times 10^{-7}$~M$_\odot$yr$^{-1}$. 
More recently, \citet{2011MNRAS.410.1761K} recomputed the orbital solution of this system and found a period of 96.4371 days. 

The revised {\it Hipparcos} parallax \citep{2007A&A...474..653V} is 3.58$\pm$0.48 mas, which leads to a linear radius of 44.7$\pm$6.0~R$_\odot$, given our diameter measurement of 1.49 mas. 
Using the same procedure as for HD~190658, 
we derive a bolometric magnitude of $-1.78$ (or a luminosity of 410~L$_\odot$) and an effective temperature of  $4000\pm100$~K (corresponding indeed to a K4--5~III star). 

The measured $v \sin i$ is compatible with the star being in synchronous rotation provided we assume the inclination $i=67\pm23^{\circ}$. In this case, we can use the observed spectroscopic mass function, $f(m)=0.13589$~M$_\odot$, to relate the mass of the hot star, $M_1$, to that of the giant, $M_2$. For $M_2 < 1.5$~M$_\odot$ (as required by the observed stellar parameters of the giant), we have $M_1 < 1$~M$_\odot$, and the Roche lobe filling factor is always larger than 90\% ($\pm$ 10\%). In fact, if the mass of the giant is below 1.12~M$_\odot$, it is filling fully its Roche lobe. This is in agreement with \citet{1990AJ....100..554H}, even though this author derived smaller values for the masses and radius. As for HD 190658, the Roche lobe filling factor is not very sensitive to the inclination that we use, i.e. varies by a few percents only. We note that the spectroscopic mass function allows the secondary to be more massive than the giant primary (that is, the system  is an Algol) only if the inclination is smaller than $60^{\circ}$ (and the primary is not very massive), which is not very likely as we see ellipsoidal variations.  

Given such a Roche lobe filling factor and the presence of ellipsoidal variations, we could imagine that the star is tidally distorted. Our visibilities seem to indicate this, as the model shown in Fig.~\ref{fig:diam1} is not a very good fit to the data. Using an elongated disc or an elongated Gaussian instead of a uniform disc, we can obtain a much better fit to the visibilities (see Figs.~\ref{fig:hr352} and \ref{fig:hr352img}). While an elongated disc can be seen as a rough approximation of a tidally distorted star, an elongated Gaussian could indicate that it is the wind that is filling the Roche lobe,  as in this latter case we may expect some density gradient. At this stage, we cannot distinguish between an elongated disc or an elongated Gaussian, as the $\chi^2$ are similar, but there is a clear indication of distortion.  The resulting reduced $\chi^2$ drops from 2.48 for a uniform disc to 1.25 for an elongated Gaussian 
with an elongation ratio of 1.16 (the star is 1.38$\times$1.6 mas wide). We defer to a further paper a detailed comparison with a modeling code, and after we have obtained additional data for this system to show how the angular diameter changes with time. However, we should mention that using the ephemeris of \citet{2011MNRAS.410.1761K}, our observations were done at a spectroscopic phase of 0.46, i.e. when the star is moving almost the fastest away from us. This would correspond to a photometric phase of 0.21 and thus close to where we expect the highest deformation of the star. The deformation we measure is oriented between $125$ and $150$ degrees -- depending on the solution we retain. We note that the data are not good enough for now to really constrain the shape of the star: it is also possible to obtain a similarly good fit when using an elongated disc with a major axis of 1.71 mas and an elongation ratio of 1.19, or when using a geometrical model that contains both an elongated disc and a point source, which would contribute about 2.7\% to the flux in the $H$-band. The errors on the parameters are too large, though, in this latter case to provide useful constraints. We note that the closure phases are all smaller than 2 degrees, and compatible with zero given the error bars (see Fig.~\ref{Fig:PhasesHD352} online). Clearly, more data are required on this system, in particular with longer baselines to reach the second lobe, as this makes it easier to confirm the elongated disc model.

\begin{figure}
\includegraphics[width=9cm]{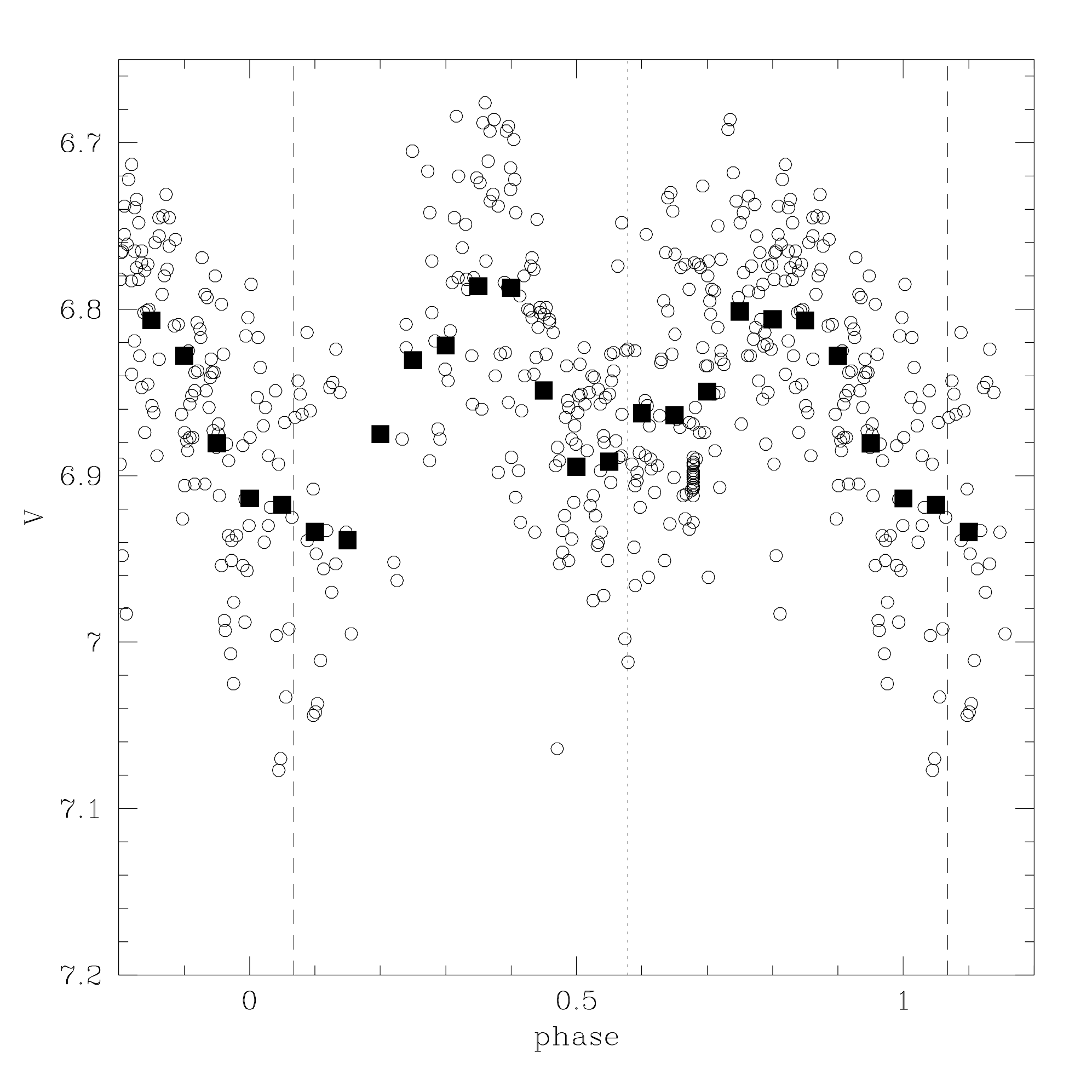}
\caption{\label{Fig:ASAS}
The ASAS $V$ lightcurve of V1261~Ori, phased with the orbital period of 638.24~d. }
\end{figure}

\begin{table}
\caption{\label{Tab:HD35155}
Revised orbital elements of V1261~Ori based 
on two new HERMES/Mercator measurements (see text).
}
\begin{tabular}{lll}
\hline
   &  1998 orbit$^a$ & Revised orbit \\ 
\hline
$P$(d) & $640.5\pm2.8$ & $638.24\pm0.28$ \\
$e$    & $0.07\pm0.03$ & $0.07\pm0.01$ \\
$T_0$ (HJD$ - 2\ts400\ts000$) & $48\,092\pm58$ & $53\,215\pm21$ \\
$\omega$ $(^\circ)$ & $232\pm33$ & $243\pm12$ \\
$K_1$ (km s$^{-1}$) & $7.88\pm0.28$ & $7.91\pm0.12$ \\
$V_0$ (km s$^{-1}$) & $79.8\pm0.2$ & $79.77\pm0.09$ \\   
$f(m)$ (M$_{\odot}$) & 0.032 & 0.032 \\
$\sigma (O-C)$ (km s$^{-1}$) & 0.8 & 0.7 \\
\hline
\end{tabular}
\tablefoot{$^a$ \citet{1998A&A...332..877J}}
\end{table}

\begin{figure}
\includegraphics[width=9cm]{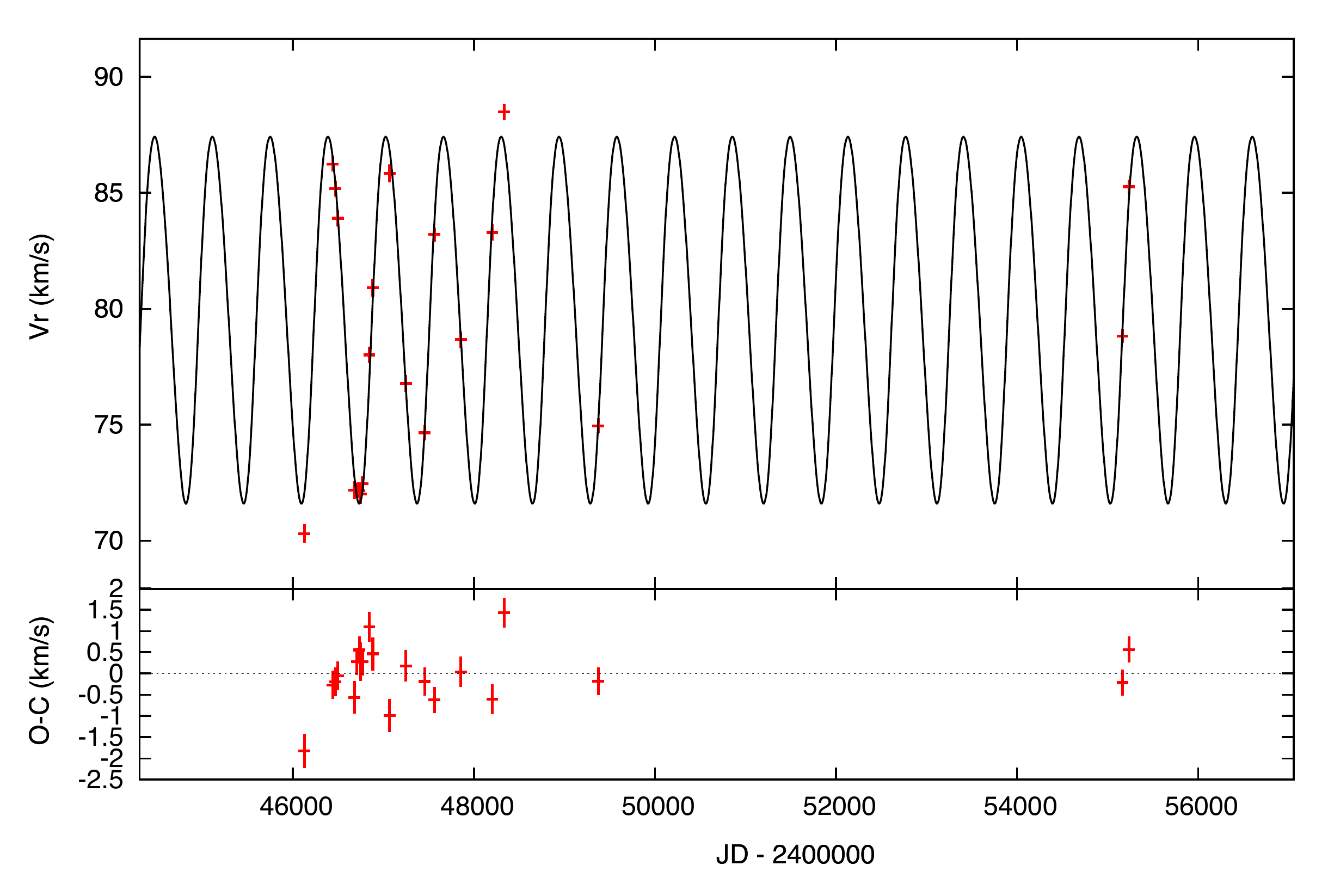}
\caption{\label{Fig:35155_VR}
The orbit of V1261~Ori, combining old CORAVEL measurements with recent HERMES/Mercator ones.
}
\end{figure}

\subsection{V1261~Ori}
V1261~Ori (HD 35155, HIP~25092) is an extrinsic S star that shows evidence for mass-transfer activity and is therefore also classified as a symbiotic star \citep{1991ApJ...383..842A,2002A&A...396..599V}. As an extrinsic S star, it is the outcome of a previous mass transfer which polluted it in carbon and s-process elements. The hot star should thus, according to this scenario,   be a white dwarf.   
\citet{2007A&A...474..653V} gives a parallax $\varpi = 3.47 \pm 0.84$ mas, putting it 288 pc away. 
\citet{Pourbaix2000}, however, obtained the astrometric orbital elements for that star
by reprocessing the {\it Hipparcos} IAD using the knowledge of the spectroscopic orbital 
elements.  A satisfactory solution (in terms of reduced $\chi^2$) emerges, with 
a revised parallax $\varpi = 1.96^{+1.44}_{-0.83}$ (i.e. a distance of 510$^{+375}_{-215}$ pc) , and an orbital 
inclination of $100^\circ\pm26^\circ$. Such a value for the orbital inclination 
possibly allows for eclipses, and one such event has been reported  by \citet{Jorissen1992} in the Str\"omgren $y$ filter. A weakening of the companion UV continuum, at the phase 
corresponding to the superior conjunction, has 
also been reported by \citet{1991ApJ...383..842A} from a series of IUE spectra. Obviously, the eclipsed
light cannot be that of the white dwarf, which is too faint, but must come
from, e.g.,  a hot spot. That spot is obviously intermittent, since the eclipsing behaviour 
is not seen in more recent years, as shown by the ASAS lightcurve \citep[Fig.~\ref{Fig:ASAS},
where it is phased with the orbital period of 638.24~d;][]{Pojmanski1997}. As already shown by 
Gromadzki, Mikolajewska \& Soszynski (\citeyear{Gromadzki2013}), there is a clear modulation with the orbital motion, 
in the form of a double sine-wave, which suggests an ellipsoidal variation. 
The large filled squares on Fig.~\ref{Fig:ASAS} correspond to the average of the data points 
over a phase range of $\pm0.05$. The 'noise' around this average curve 
is caused by semi-regular variations with a time scale of the order of 56~d and a semi-amplitude 
of 0.1~mag. The dashed and 
dotted lines on Fig.~\ref{Fig:ASAS} correspond to the superior (companion behind the S star) 
and inferior conjunctions, respectively,
based on updated orbital elements, thanks to two recent HERMES/Mercator radial-velocity measurements
obtained on HJD~2455164.69083, with $Vr = 78.826\pm0.006$~km s$^{-1}$, and 
HJD~2455235.446 with $Vr = 85.265\pm0.008$~km s$^{-1}$ \citep{Jorissen1992b}; these velocities are on the IAU system
to ensure consistency with the older CORAVEL velocities and allowed 
us to accurately constrain the orbital period, as shown on Table~\ref{Tab:HD35155} and 
Fig.~\ref{Fig:35155_VR}. The eccentricity as well is accurately determined thanks to
these two recent velocity measurements, and happens to be small, albeit significantly different  
from zero. The combination of a non-zero eccentricity with ellipsoidal variations is a bit 
surprising, since ellipsoidal variations imply a filling factor close to unity, 
which rapidly (i.e., in a few $10^7$~yr) circularises the orbit and synchronises as well 
the giant's spin with the orbital motion. 

To add to the puzzle, our radius measurement does not lead to a filling factor close to unity in fact, as
we now discuss. The semi-major axis of the 
photocentric orbit around the centre of mass of the system (this orbit is
equivalent to the orbit of the giant around the centre of mass if the white 
dwarf companion contributes no light to the system)   
is found to be   $1.84\pm0.9$~mas \citep{Pourbaix2000}.   Since \citet{Jorissen1992} estimate 
the mass ratio to be $M_1/M_2 = 3$ \citep[based on the ratio of the 
velocity-curve 
amplitudes, as lines from the hot companion are visible in the IUE spectrum; ][]
{1991ApJ...383..842A}, the relative orbit must have a semi-major 
axis of $4 \times 1.84 = 7.36$~mas, with a fractional Roche-lobe size of 0.476
around the giant, or 3.5$\pm1.7$~mas, to be compared with the 
measured radius of 1.12~mas (Table~1). The red giant in the V1261~Ori  system 
thus lies well inside its Roche lobe, with a fractional Roche radius of 0.32$\pm0.16$,
independent of any assumption on the distance. The interaction in this system 
is thus most likely due to wind mass transfer. The ellipsoidal variations 
reported by Gromadzki et al. (2013, and Fig.~\ref{Fig:ASAS})
are thus puzzling.  Moreover, ellipsoidal variations are generally concomitant with synchronisation 
between rotation and orbital revolution. 
From the measured diameter of 2.25~mas (Table~1) and the parallax of 
Pourbaix \& Jorissen (2000), a linear radius
of $123^{+90}_{-52}$~R$_\odot$ is derived, implying a rotation velocity of 5.6 to 16.8~km s$^{-1}$
if synchronised with the orbital motion.
The width of the HERMES cross-correlation
function (CCF; 4.2 km s$^{-1}$) is typical of M~giants \citet{Famaey2009}. 
For this system seen almost edge-on, the measured CCF width is thus not in favour of 
the giant being synchronised with the orbital motion.
The interpretation of the orbital modulation of the photometric variations as being due 
to ellipsoidal variations, despite seeming likely from the shape of the light variations, 
is supported neither by the small filling factor, nor by the absence of synchronisation. 
It thus remains an unresolved puzzle for this system in particular. One possibility could be
tidally-enhanced pulsations, although, again, the relatively low Roche-lobe filling would argue against this.

Finally, from our diameter measurement and the estimate of $BC_K$ (in agreement with the value quoted by  \citealt{1998A&A...329..971V}), we 
derive a distance-independent effective temperature of $3470\pm 60$ K, leading to $M_{\rm bol}=-3.5\pm1.2$. 


\begin{table*}[ht]
   \centering
    \caption{Result of fitting the parameters for AG Peg, for three assumed values of the inclination $i$, in the range 30 to 90 degrees. In each case, we indicate the mass of the giant, $M_2$, of its hot companion, $M_1$, the semi-major axis, $a$ (in AU), the bolometric magnitude, the Roche lobe radius, $R_L$ (in R$_\odot$), the radius of the giant, $R_2$ (in R$_\odot$), the filling factor, $f=R_2/R_L$, the rotational velocity, $v \sin i$ (in km s$^{-1}$), assuming rotation is coplanar to the orbital motion, and the resulting parallax, $\varpi$, in mas.}
   \label{tab:agpeg}
   \begin{tabular}{lcccccccc} 
   \hline
\multicolumn{9}{c}{ $i = 30^{\circ}$}\\

\hline\hline
	   $M_2$&	    $M_1$&	     $a$&	  M$_{\rm bol}$&	    $R_L$&	   $R_2$&	     $f=R_2/R_L$&	     $ v \sin i$&	   $\varpi$ (mas) \\
\hline
	   0.8&	0.5944&	  1.91&	-1.499&	 166.6&	 47.51&	0.2851&	 1.481&	 2.299\\
	     1.0&	0.6704&	 2.029&	-1.912&	 180.9&	 57.45&	0.3176&	 1.791&	 1.901\\
	   1.3&	0.7746&	 2.181&	-2.375&	 199.2&	 71.11&	0.3569&	 2.217&	 1.536\\
	   1.6&	0.8702&	 2.311&	-2.728&	 215.1&	 83.64&	0.3889&	 2.607&	 1.306\\
	     2.0&	0.9879&	 2.462&	-3.094&	 233.6&	 99.01&	0.4238&	 3.087&	 1.103\\
	   2.5&	 1.123&	 2.626&	-3.449&	 253.8&	 116.6&	0.4596&	 3.635&	0.9366\\
	     3.0&	 1.249&	 2.769&	-3.733&	 271.5&	 132.9&	0.4895&	 4.143&	0.8219\\
	   3.5&	 1.368&	 2.897&	-3.969&	 287.6&	 148.2&	0.5152&	 4.618&	0.7373\\
	     4.0&	  1.48&	 3.014&	-4.171&	 302.2&	 162.6&	 0.538&	 5.068&	0.6719\\
\hline
\multicolumn{9}{c}{$i = 50^{\circ}$}\\
\hline\hline
	   $M_2$&	    $M_1$&	     $a$&	  M$_{\rm bol}$&	    $R_L$&	   $R_2$&	     $f=R_2/R_L$&	     $ v \sin i$&	   $\varpi$ (mas) \\
\hline
	   0.8&	 0.339&	 1.785&	-1.499&	 174.5&	 47.51&	0.2723&	 2.269&	 2.299\\
	     1.0&	0.3865&	 1.906&	-1.912&	 189.6&	 57.45&	 0.303&	 2.744&	 1.901\\
	   1.3&	0.4517&	 2.061&	-2.375&	 209.1&	 71.11&	0.3402&	 3.396&	 1.536\\
	   1.6&	0.5116&	 2.193&	-2.728&	 225.9&	 83.64&	0.3703&	 3.995&	 1.306\\
	     2.0&	0.5855&	 2.347&	-3.094&	 245.5&	 99.01&	0.4032&	 4.729&	 1.103\\
	   2.5&	0.6709&	 2.512&	-3.449&	 266.9&	 116.6&	0.4369&	  5.57&	0.9366\\
	     3.0&	0.7503&	 2.656&	-3.733&	 285.8&	 132.9&	0.4651&	 6.348&	0.8219\\
	   3.5&	0.8251&	 2.786&	-3.969&	 302.7&	 148.2&	0.4894&	 7.076&	0.7373\\
	     4.0&	0.8963&	 2.903&	-4.171&	 318.3&	 162.6&	0.5108&	 7.765&	0.6719\\
\hline
\multicolumn{9}{c}{$i = 90^{\circ}$}\\
\hline\hline
	   $M_2$&	    $M_1$&	     $a$&	  M$_{\rm bol}$&	    $R_L$&	   $R_2$&	     $f=R_2/R_L$&	     $ v \sin i$&	   $\varpi$ (mas) \\
\hline
	   0.8&	0.2452&	 1.735&	-1.499&	 180.1&	 47.51&	0.2638&	 2.962&	 2.299\\
	     1.0&	0.2808&	 1.857&	-1.912&	 195.7&	 57.45&	0.2935&	 3.581&	 1.901\\
	   1.3&	0.3298&	 2.012&	-2.375&	 215.9&	 71.11&	0.3294&	 4.434&	 1.536\\
	   1.6&	0.3748&	 2.145&	-2.728&	 233.4&	 83.64&	0.3584&	 5.215&	 1.306\\
	     2.0&	0.4304&	 2.299&	-3.094&	 253.7&	 99.01&	0.3902&	 6.173&	 1.103\\
	   2.5&	0.4947&	 2.464&	-3.449&	 275.9&	 116.6&	0.4227&	 7.271&	0.9366\\
	     3.0&	0.5546&	 2.609&	-3.733&	 295.4&	 132.9&	0.4499&	 8.286&	0.8219\\
	   3.5&	0.6111&	 2.739&	-3.969&	   313&	 148.2&	0.4733&	 9.237&	0.7373\\
	     4.0&	0.6648&	 2.857&	-4.171&	 329.1&	 162.6&	 0.494&	 10.14&	0.6719\\
\hline
   \end{tabular}
  \end{table*}

\subsection{FG Ser}
A bona-fide symbiotic star, FG Ser (AS 296) is listed as M3e in {\it SIMBAD}, while \citet{1999A&AS..137..473M} give it a M5 spectral type.
\citet{2000AJ....120.3255F} found a revised orbital period of 633.5$\pm$2.4 days, and an epoch of conjunction $T_0=$~JD~$2,451,031.4\pm2.9$. 
\citet{2007MNRAS.380.1053Z} measure a $v \sin i =9.8\pm1$ km s$^{-1}$, and assume a radius $R=139.6$~R$_\odot$, from
which they deduce that the system is synchronised within the measurement errors. They also assume a mass of the giant of 1.7~M$_\odot$. 

We measure an angular diameter for FG~Ser that changes with the orbital phase: $0.83\pm0.03$ mas and  $0.94\pm0.05$ mas, 41 days later. 
According to the above mentioned ephemeris, this would correspond to orbital phases very close to the conjunction, and to a difference in phase of 0.07 only. 

As no distance has been estimated for this object, it is impossible to directly determine the linear radius and compute the Roche lobe filling factor. We can use some indirect method, however.
Using the distance-independent method and the mean of our angular diameters we determine the temperature of the red giant: T$_{\rm eff}=3100\pm100$~K -- a value not atypical for a M5 giant. We can now also assume that the giant is on the first ascent giant branch and use the relation between the radius and the luminosity of the star, as a function of its mass;  \citet{2000MNRAS.315..543H} provide a fit to such a relation for solar abundances:
\begin{equation}
R_2 = 1.1 M_2^{-0.3}\left( L^{0.4} + 0.383 L^{0.76} \right).
\end{equation}

We can further use the relation $L=R_2^2 \theta^4$, with $\theta=T_{\rm eff}/5777$, as well as the spectroscopic mass function $f(m)=0.0218\pm0.0025$~M$_\odot$ from \citet{2007MNRAS.380.1053Z} to relate $M_1$ to $M_2$ for a given inclination. This then allows us to compute the final Roche lobe radius and compare it to the radius of the star. We find that in all cases, the giant is filling its Roche lobe.
For example, for $i=90^{\circ}$, assuming a giant mass of $M_2=1$~M$_\odot$ leads to a companion mass of $M_1=0.34$~M$_\odot$, a giant radius of 158~R$_\odot$, to be compared to a Roche lobe radius of 162~R$_\odot$. Changing $i$ to $50^{\circ}$ leads to  $M_1=0.47$~M$_\odot$ and a Roche lobe radius of 157~R$_\odot$. Increasing the mass of the giant further strengthens this conclusion.
Assuming a 1~M$_\odot$ giant would also lead to a synchronised rotational velocity close to 10 km s$^{-1}$, as observed. It would imply a distance of the system of 1700 pc ($\varpi = 0.59$ mas, well beyond the capabilities of {\it Hipparcos}).

We conclude that FG Ser is a nominal case of a synchronous giant filling its Roche lobe. The fact that we measure a different angular diameter depending on the orbital phase -- albeit with a very low significance only -- confirms the nature of the ellipsoidal variations and the fact that the inclination has to be close enough to an edge-on system. Here again, a detailed modelling of the combined interferometric and photometric data is outside the scope of this paper.

\subsection{AG Peg}
The symbiotic binary AG Peg (HD 207757, HIP 107848) is unique among the objects in our sample as it is known to have undergone a slow nova eruption when it rose from 9th to 6th magnitude, starting in the mid-1850s. The hot companion is therefore unambiguously a white dwarf, while the red giant primary, which has an M2--M3 spectral type, does not apparently fill its Roche lobe, and loses mass through a low-velocity wind  (\citealt{2001AJ....122..349K}, and references therein). Given the low luminosity of the hot component and very slow outburst development, the white dwarf in AG Peg is likely of low mass, and indeed the radial velocity curve hints at a mass $M_1=0.46\pm0.10$~M$_\odot$ (Mikolajewska, 2010).

\citet{2000AJ....119.1375F} provide the orbital elements of this binary system: $P=818.2\pm1.6$~d, $e=0.11\pm0.039$, $f(m)=0.0135$~M$_\odot$. {\it Hipparcos} found the system to present large photometric variations ($A=0.47$ mag) which fit the orbital period, while \citet{2007BaltA..16...49R} claim that the light curve of AG Peg does not show any evidence for a tidally distorted giant. Unfortunately, the {\it Hipparcos} astrometry does not provide a significant value for the parallax, which is therefore most likely below 1 mas. 

\citet{2007MNRAS.380.1053Z} report a $v \sin i = 8.5\pm1.5$ km s$^{-1}$ and a M4~III spectral type and radius of 71.5~R$_\odot$ for the giant. 
We find an angular diameter of 1 mas for the giant in this system, which when combined to the $J$ and $K$ magnitudes in the same way as before, leads us to an effective temperature of  3550$\pm$120 K, in agreement with the value we would derive from the $(V-K)$ index and the relations of \citet{1980ApJ...235..126R} or \citet{2008ApJSVanBelle} and in agreement with the spectral type.  We can then again make use of these parameters, as well as of the spectroscopic orbital elements and a radius-luminosity relation, to derive the parameters of the system, depending on the inclination (see Table~\ref{tab:agpeg}). Taken at face values, the rotational velocity and the assumption of synchronism, as well as the value of $M_1$ quoted above, would imply an inclination close to $90^{\circ}$, $M_2\sim2-3$~M$_\odot$, $R_2\sim99-133$~R$_\odot$ and a Roche lobe filling factor $f\sim0.39-0.45$. In any case, the filling factor lies -- for all reasonable values of the system's parameters -- in the range 0.26 to 0.54, but the lowest values can be discarded as we need to have a parallax small enough not to have been measurable by {\it Hipparcos}, while the higher values would imply a white dwarf mass which is too high.

\subsection{ER~Del}
This is another example of an S star (S5.5) that exhibits symbiotic-like features \citep{1979ApJ...234..538A}. From its $(V-K)$, we can estimate a temperature around 3500 K using the relation of \citet{1980ApJ...235..126R}, while the $(J-K)$ can be used to compute BC$_K$=2.9. 
Using our angular diameter of 0.61 mas and the above derived BC$_K$, we can also determine the effective temperature and find $T_{\rm eff}=3470\pm$ 160~K.

\begin{table}
\caption{
Radial velocities of ER~Del.
}
\begin{tabular}{rrrr}
\hline\hline
JD~~~~~~ & {$Vr$}~~~ & error~& Inst~~~\\
$-2\ts400\ts000$ & (km s$^{-1}$)& (km s$^{-1}$)&\\
\hline
48452.809 &-58.97 &0.35 & COR\\
48841.707 &-51.57 &0.27 & COR\\
49522.829 &-43.28 &0.31 & COR\\
50353.367 &-57.30 &0.48 & COR\\
50379.288 &-56.76 &0.27 & COR\\
50622.500 &-55.10 &0.30 & COR\\
55001.683 & -52.575 &0.008 & HER\\
55024.656 & -53.886 &0.013 & HER\\
55051.569 & -53.280 &0.014 & HER\\
55098.444 & -53.298 &0.008 & HER\\
55350.725 & -49.102 &0.011 & HER\\
55417.574 & -45.855 &0.009 & HER\\
55503.372 & -45.077 &0.006 & HER\\
55718.701 & -44.025 &0.012 & HER\\
55834.507 & -44.386 &0.006 & HER\\
56032.739 & -43.705 &0.009 & HER\\
56062.644 & -43.443 &0.030 & HER\\
56102.699 & -42.983 &0.011 & HER\\
56153.572 &-44.102  &0.013 & HER\\
56464.724 &-52.208  &0.016 & HER\\
56486.683 &-51.657 & 0.011 & HER\\
56535.526 &-54.601 & 0.016 & HER\\
\hline
\end{tabular}
\label{Tab:VRERDel}
\end{table}

\begin{table}
\caption{
Orbital elements of ER Del.
}
\begin{tabular}{lll}
\hline\hline
& Value & error \\
\hline
$P$ (d) & 2094.2 & 4.4 \\
$e$     & 0.228 & 0.016 \\
$\omega$ ($^\circ$) & 127.8 & 3.6 \\
$V_0$ (km s$^{-1}$) & $-49.37$ & 0.07 \\
$K_1$ (km s$^{-1}$) & 7.27 & 0.01 \\
$T$ (JD) & 2\ts454\ts465.3 & 19.4 \\
$a_1 \sin i$ (AU) & 1.36 & 0.01\\ 
$f(m)$ (M$_{\odot}$) & 0.077 & 0.001\\
\hline
\end{tabular}
\label{Tab:ERDelorbit}
\end{table}

\begin{figure}
\includegraphics[width=9cm]{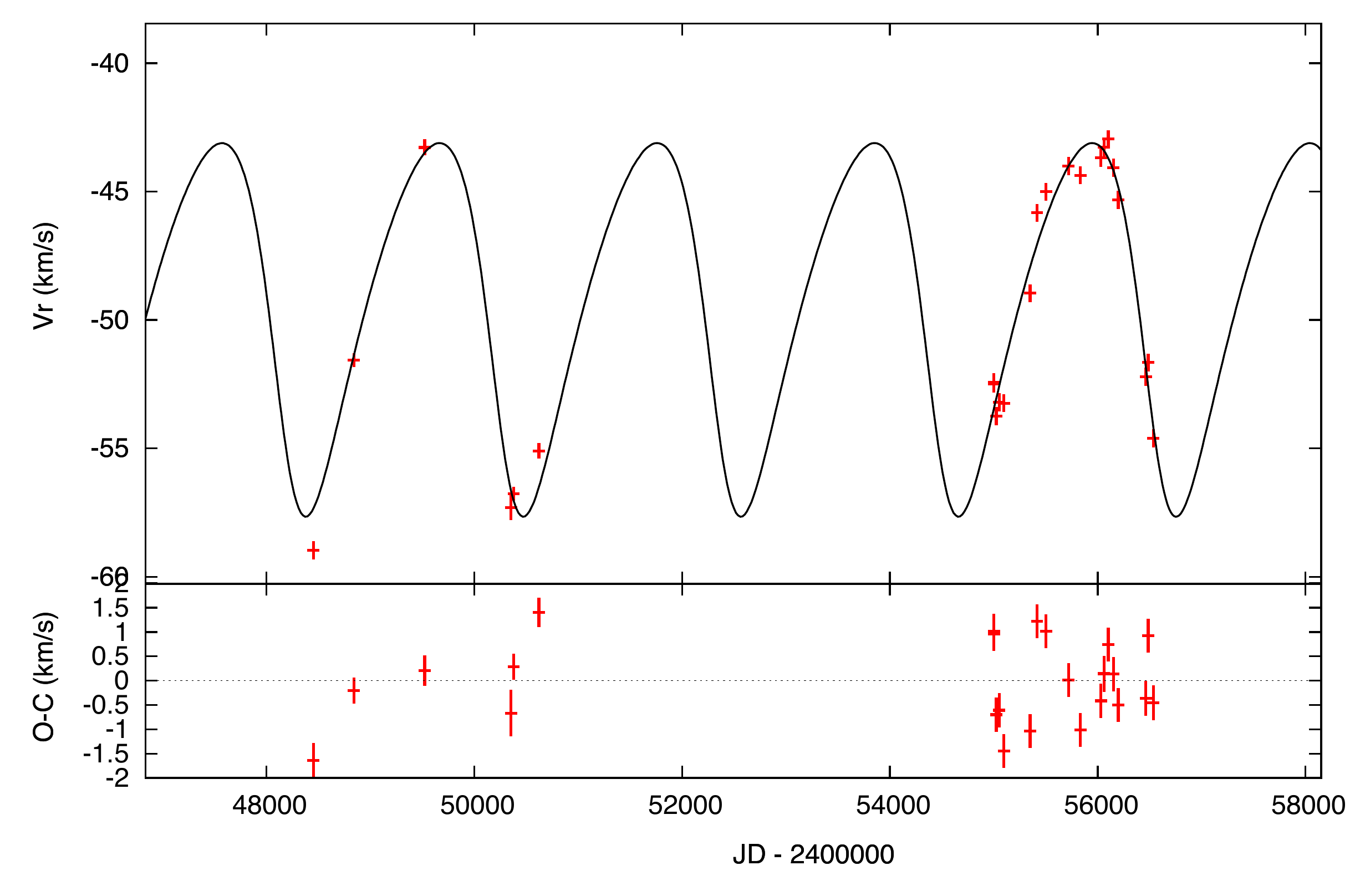}
\caption{
The orbital solution for ER~Del. The lower panel provides the O-C residuals. The
error bars of the HERMES data points have been artificially increased to 0.3~km s$^{-1}$.}
\label{Fig:ERDelorbit}
\end{figure}

Radial-velocity observations were obtained with CORAVEL \citep[COR;][]{Baranne1979} in 1991-1997, and then from 2009
onwards with the HERMES/Mercator spectrograph \citep[HER;][]{Raskin2011}. They are on the system of IAU radial-velocity 
standard stars as defined by \citet{Udry1999} and displayed in Table~\ref{Tab:VRERDel}.
The orbital solution is shown in Table~\ref{Tab:ERDelorbit} and Fig.~\ref{Fig:ERDelorbit}.
It has large residuals ($<|O-C|> = 0.71$~km s$^{-1}$), as is usual for M and S stars 
\citep[see Fig.~1 of][and Fig.~2 of Famaey et al. 2009]{1998A&A...332..877J}\nocite{Famaey2009}.
Also here, the width of the HERMES CCF is typical of M giants and we cannot measure the rotational velocity of this giant.

By using the same methodology as above, we are led to conclude that the filling factor is between 0.2--0.3, for a wide range of parameters. If we assume synchronisation, the derived radius leads to rotation velocities of 2--3 km s$^{-1}$, which would indeed be below the detection limit.
If we make use of the fact that ER Del is an S star and so assume the typical masses for the components of such a post-mass transfer system, and in particular for the white dwarf, we are led to the conclusion that $i > 40^\circ$. For $i=50^\circ$ and $M_1=0.7$~M$_\odot$, we find $M_2=2$~M$_\odot$, $M_{\rm bol}=-3.3, d=1750$~pc, and $\varpi=0.58$~mas. For $i=90^\circ$ and $M_1=0.7$~M$_\odot$, we find $M_2=3.25$~M$_\odot$, $M_{\rm bol}=-4, d=2600$~pc, and $\varpi=0.39$~mas. In any case, the giant is well within its Roche lobe.

   \begin{table*}[htbp]
   \centering
      \caption{Summary table of parameters of our target stars. These are ranked by orbital period. $e$ is the eccentricity, $f(m)$ the binary spectroscopic mass function.}
   \begin{tabular}{llrcllrrccc} 
\hline\hline
      Star  		& Parallax 	& Period 		& e & $f(m)$ & $T_{\rm eff}$	& Radius         & M$_{\rm bol}$ 	&	$f=R/R_L$  & $v\sin i$ & synchronised?\\
               		& (mas)        	& (days)      	&&	(M$_\odot$) & (K)			      & (R$_\odot$) & 					&			& km s$^{-1}$ &	\\
      \hline
HD 352 		&    3.58     	&96.4 & 0.022 &0.1359 &   	4000$\pm$100	& 	44.7$\pm$6.0	& 		$-1.78$			&	0.8--1 & 22 & y\\
HD 190658 	&   2.4		&198.7& 0.05 & 0.045 & 	     3263$\pm$35   &     104$\pm$56    &			$-2.6$		&   0.4--1 & ? & --\\
FG Ser 		& $<0.6$		&633.5 & 0.0 &   0.0218 &   3100$\pm$100   &     $\sim 160$  	&			$<-3.6$		& 	$\sim 1$ & 9.8 & y		\\
V1261~Ori  	& $\sim1.96$	&638.2  	&0.07 & 0.032 &  	3470$\pm$60	& 	 $120^{+90}_{-52}$	& 		$-3.5$			&	 $\sim$0.3 & $<$ 4 & n\\
AG Peg 		& $<1$		&819.2 &0.11 &  0.0135 & 3550$\pm$120	& 	47--163	 &		$<-1.5$			& 	0.25--0.55 & 8.5 & y\\
ER Del 		& $>0.4$		&2094.2&0.228& 0.077 &	3500$\pm$160	& 	$>115$	& 			$<-3.3$		&	$0.2-0.3$ & $<$ 4 & --\\

      \hline
   \end{tabular}
   \label{tab:radius}
\end{table*}

\begin{figure}
   \centering
\includegraphics[width=9cm]{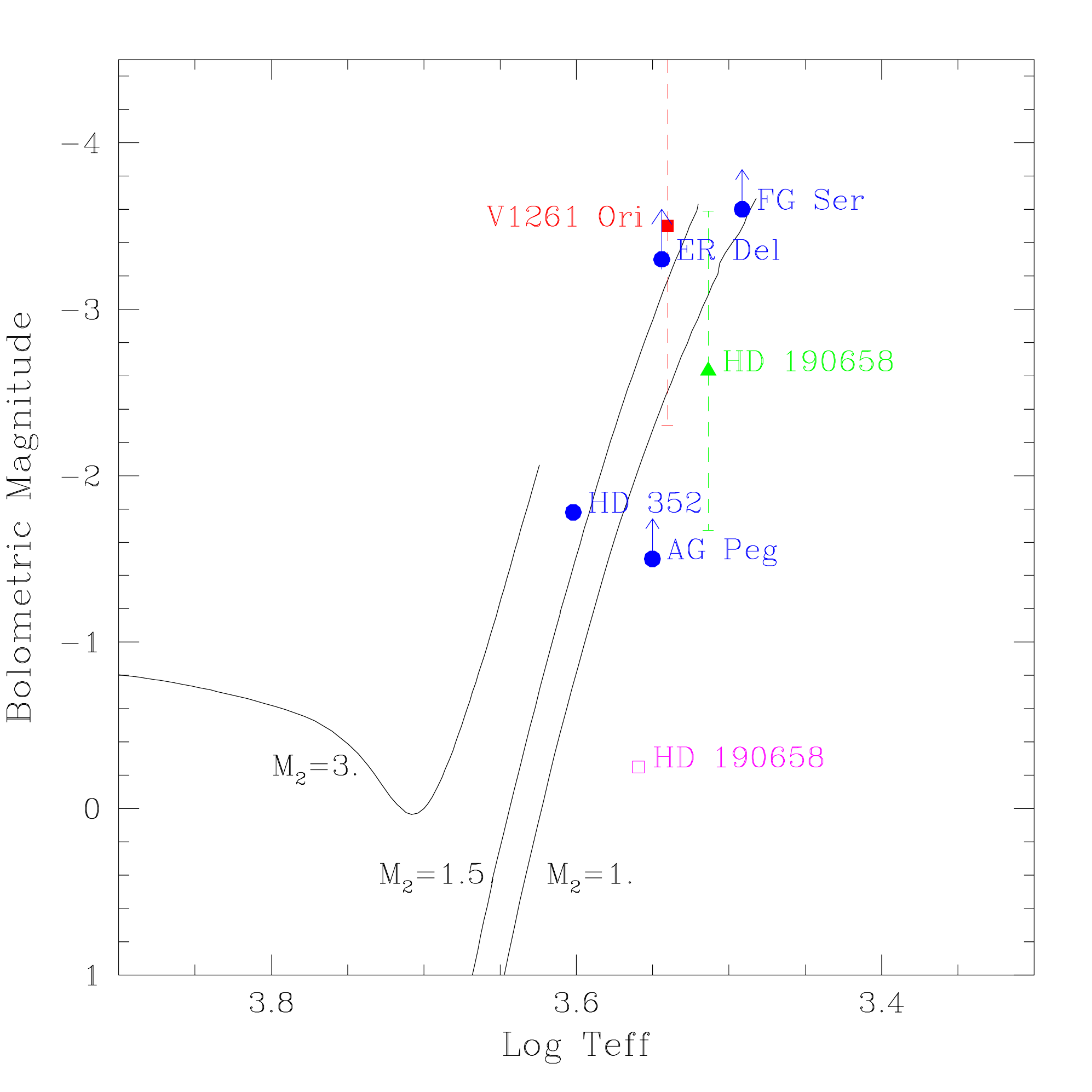}
      \caption{Hertzsprung-Russell diagramme showing the positions of our target stars (full dots, in blue), together with Yonsei-Yale (Y$^2$) evolutionary tracks for solar abundance stars with initial masses of 1, 1.5 and 3~M$_\odot$. For FG Ser, ER Del and AG Peg, we have only lower limits for the bolometric magnitude. The range for V1261 Ori is indicated with the red dashed line. For HD~190658, we indicate in magenta (open square) the value using the published  and seemingly erroneous parallax (7.92 mas)  and in green (full triangle) the value with our newly derived parallax.
              }
         \label{fig:hrdiag}
   \end{figure}

\section{Conclusions}

Table~\ref{tab:radius} summarises our results and show for each target the effective temperature, radius, bolometric magnitude and Roche lobe filling factor we are able to derive. The two last columns indicate the rotational velocity (if known) and whether the rotation is synchronised with the orbital motion. Our analysis has shown that, as already known, red giants in symbiotic systems are rather normal and obey similar relations between colour, spectral type, temperature, luminosity and radius -- a result which stems, e.g. from the distance-independent effective temperature and from the bolometric magnitude we derive. We have demonstrated that in two cases, HD 352 and FG~Ser, the star is almost or fully filling its Roche lobe, in agreement with these objects presenting ellipsoidal variations. HD~352 presents a clear deviation from a spherical shape, with an elongation of 16\% or more, while two separate observations of FG~Ser clearly indicate different angular diameters, indicative of a tidal deformation. For HD~190658, if we adopt the published parallax, we would find a Roche lobe filling factor well below 0.5, putting some doubt on the reported ellipsoidal variations. However, we have reasons to think that this parallax is in error by a factor 3, thereby implying that the star is filling between 50 and 100\% of its Roche lobe. For ER~Del, we provide the first orbit of this system, indicating that it is a rather long-period symbiotic system. ER~Del and the two other stars have Roche lobe filling factors much smaller -- between 25 and 60\%.  
It is noteworthy that V1261 Ori apparently exhibits ellipsoidal variations 
despite a filling factor of only 0.3 -- 0.5.

We also present in Fig.~\ref{fig:hrdiag} an HR-diagramme that summarises our results. Indicated are the values we derived for the effective temperatures and bolometric magnitudes  of our target stars, together with Y$^2$ evolutionary tracks \citep{2001ApJS..136..417Y} for stars with solar abundance and initial masses between 1 and 3~M$_\odot$.

For those stars for which {\it Hipparcos} provided a useful distance, the rather large error ($> 15\%$) is what limits the precision we have on the linear radius of the giant. It would thus be very useful if these distances could be determined with higher precision. It would also be as important to estimate the distances for those objects analysed here but for which no distance is currently known.  In this respect, the recent launch of the ESA GAIA satellite offers great prospects. GAIA will allow us to obtain parallaxes for our targets with a precision of a few microarcseconds \citep{2008IAUS..248..217L}, while at the same time being able to clearly disentangle the binary motion from the parallactic one \citep{2008IAUS..248...59P}, and the main source of uncertainty on the Roche-lobe filling factor will be the error arising from the interferometric measurements.
While GAIA will be able to perform such measurements for most or all symbiotic stars known, interferometric measurements are currently limited to only an handful of additional systems, those which are bright enough in the $H-$band and that have reliable orbital elements for further analysis.

\begin{acknowledgements}
It is a pleasure to thank Dimitri Pourbaix for reprocessing the {\it Hipparcos} data of HD~190658.
      Based on observations made with the ESO Very Large Telescope Interferometer under Prog. ID 089.D-0527(A).
   PIONIER is funded by the Universit\'e Joseph Fourier (UJF,
Grenoble) through its P\^oles TUNES and SMING and the vice-president of research, the Institut de Plan\'etologie et d'Astrophysique de Grenoble, the ``Agence Nationale pour la Recherche'' with the programme ANR EXOZODI, and the Institut
National des Sciences de l'Univers (INSU) with the programmes ``Programme National de Physique Stellaire'' and ``Programme National de Plan\'etologie''.
The integrated optics beam combiner is the result of a collaboration between
IPAG and CEA-LETI based on CNES R\&T funding. 
Based on observations made with the Mercator Telescope, operated on the island of La Palma by the Flemish Community, at the Spanish Observatorio del Roque de los Muchachos of the Instituto de Astrof\'\i sica de Canarias. Based on observations obtained with the HERMES spectrograph, which is supported by the Fund for Scientific Research of Flanders (FWO), Belgium , the Research Council of KU Leuven, Belgium, the Fonds National de la Recherche Scientifique (FNRS), Belgium, the Royal Observatory of Belgium, the Observatoire de Gen\`eve, Switzerland and the Th\"uringer Landessternwarte Tautenburg, Germany.
This research has made use of the Jean-Marie Mariotti Center {\tt Aspro2, LITpro} and {\tt SearchCal} 
service, available at \url{http://www.jmmc.fr/aspro}, and of the SIMBAD database,
operated at Centre de Donn\'ees astronomiques de Strasbourg (CDS), France, as well as the Smithsonian/NASA's Astrophysics Data System (ADS). JM is supported by Polish NSC
grant  DEC-2011/01/B/ST9/06145.
\end{acknowledgements}

\break

\Online

 
\begin{figure}[h]
\includegraphics[width=9cm]{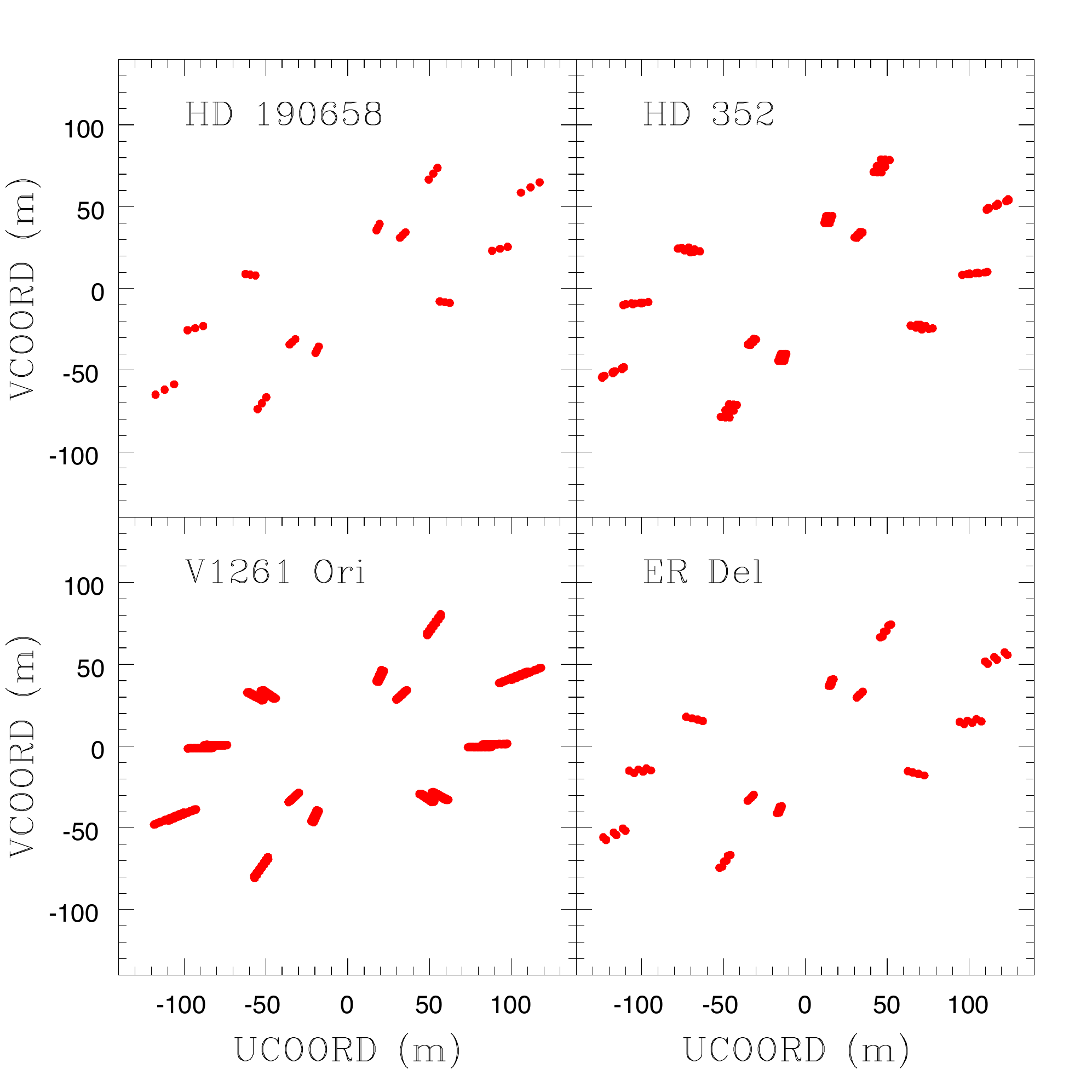}
\caption{
UV-plane coverage of our PIONIER observations for the stars HD 190658, HD 352, V1261 Ori, and ER Del. }\label{Fig:uvplane1}
\end{figure}

\begin{figure}[h]
\includegraphics[width=9cm]{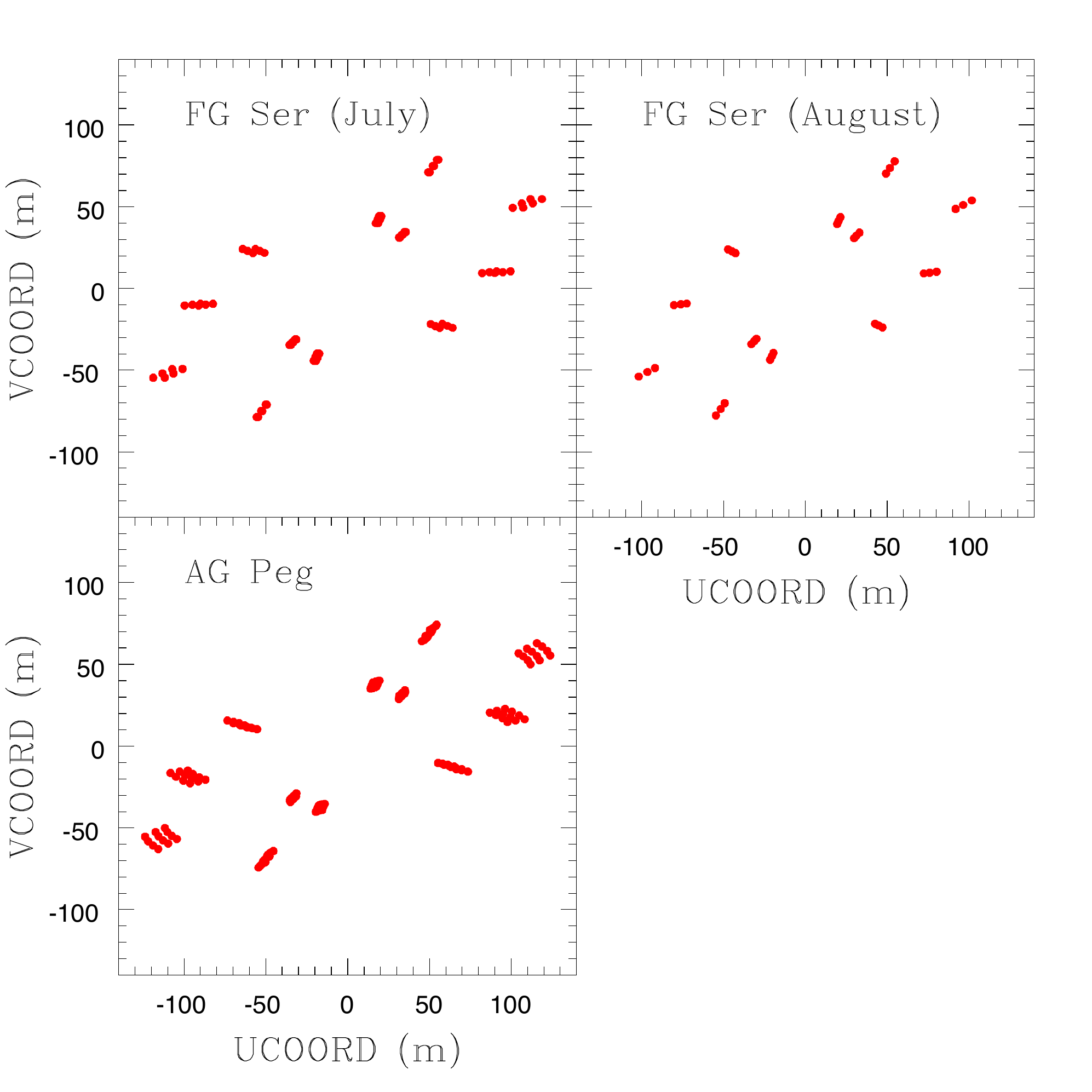}
\caption{
Same as Fig.~\ref{Fig:uvplane1} for the stars FG Ser (on 3 July and 13 August 2012) and AG Peg. }\label{Fig:uvplane2}
\end{figure}

\begin{figure}
\includegraphics[width=9cm]{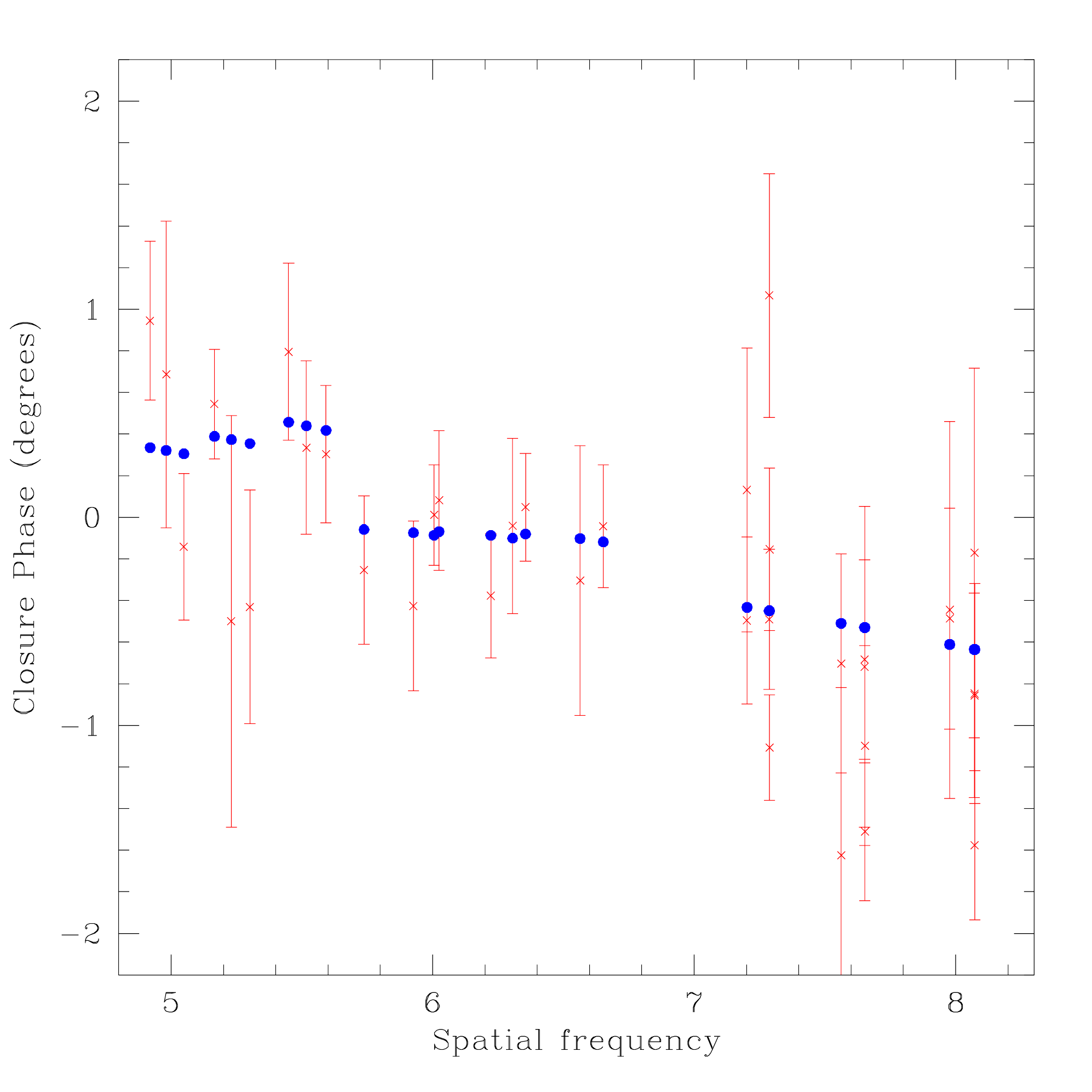}
\caption{
PIONIER closure phases for HD 352 (in degrees) as a function of the spatial frequency in 1/rad ($\times 10^7$). The data points are shown in red with error bars, while the model of an elongated disc with an additional punctual source that provides the best fit is indicated with blue solid dots. Given the error bars, the significance of these data, compared to zero closure phases, is very low, as all but three points do not deviate from zero more than the error bar.}\label{Fig:PhasesHD352}
\end{figure}

\begin{figure}
\includegraphics[width=9cm]{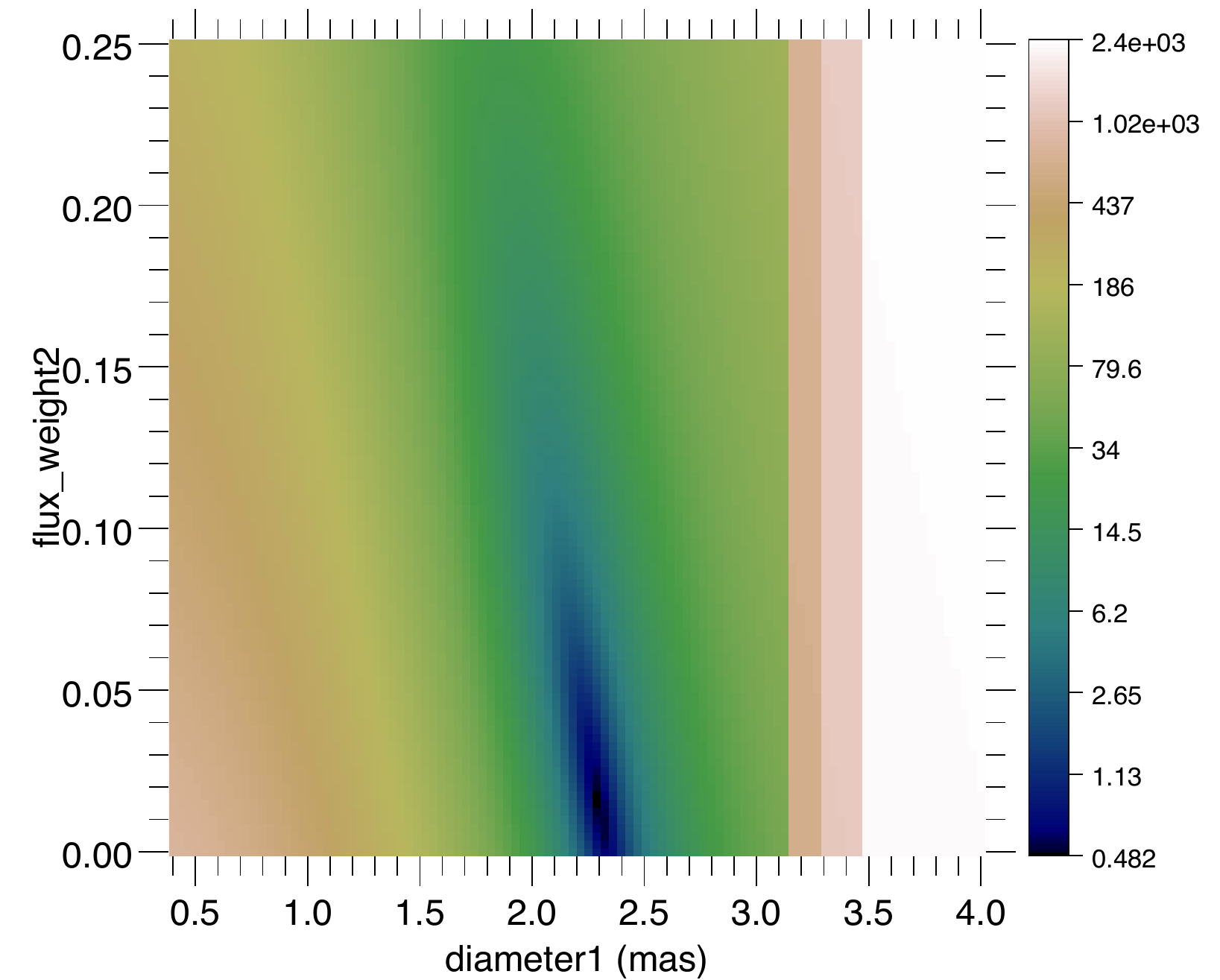}
\caption{\label{Fig:chi2map190}
Reduced $\chi^2$ maps for HD 190658: shown is the reduced $\chi^2$ (logarithmic scale) as a function of the angular diameter ($x$-axis) and the relative flux level of the background ($y$-axis) based on our PIONIER data. These maps were computed using {\tt LITpro} assuming a uniform disc plus background model. It appears clearly that the minimal reduced $\chi^2$ corresponds to the model without any added background.}
\end{figure}

\begin{figure}
\includegraphics[width=9cm]{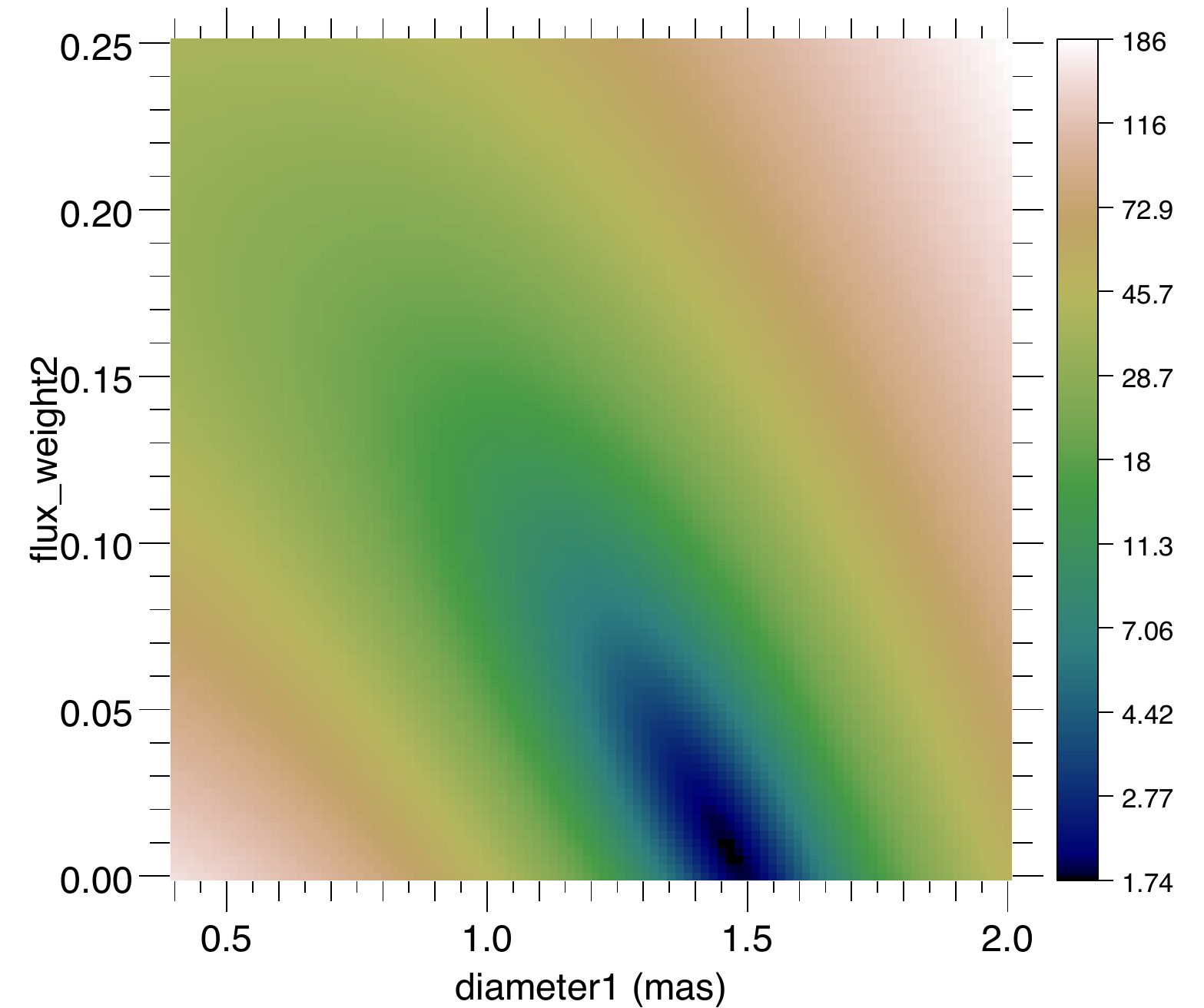}
\caption{\label{Fig:chi2map352}
Same as Fig.~\ref{Fig:chi2map190} for HD~352.}
\end{figure}

\begin{figure}
\includegraphics[width=9cm]{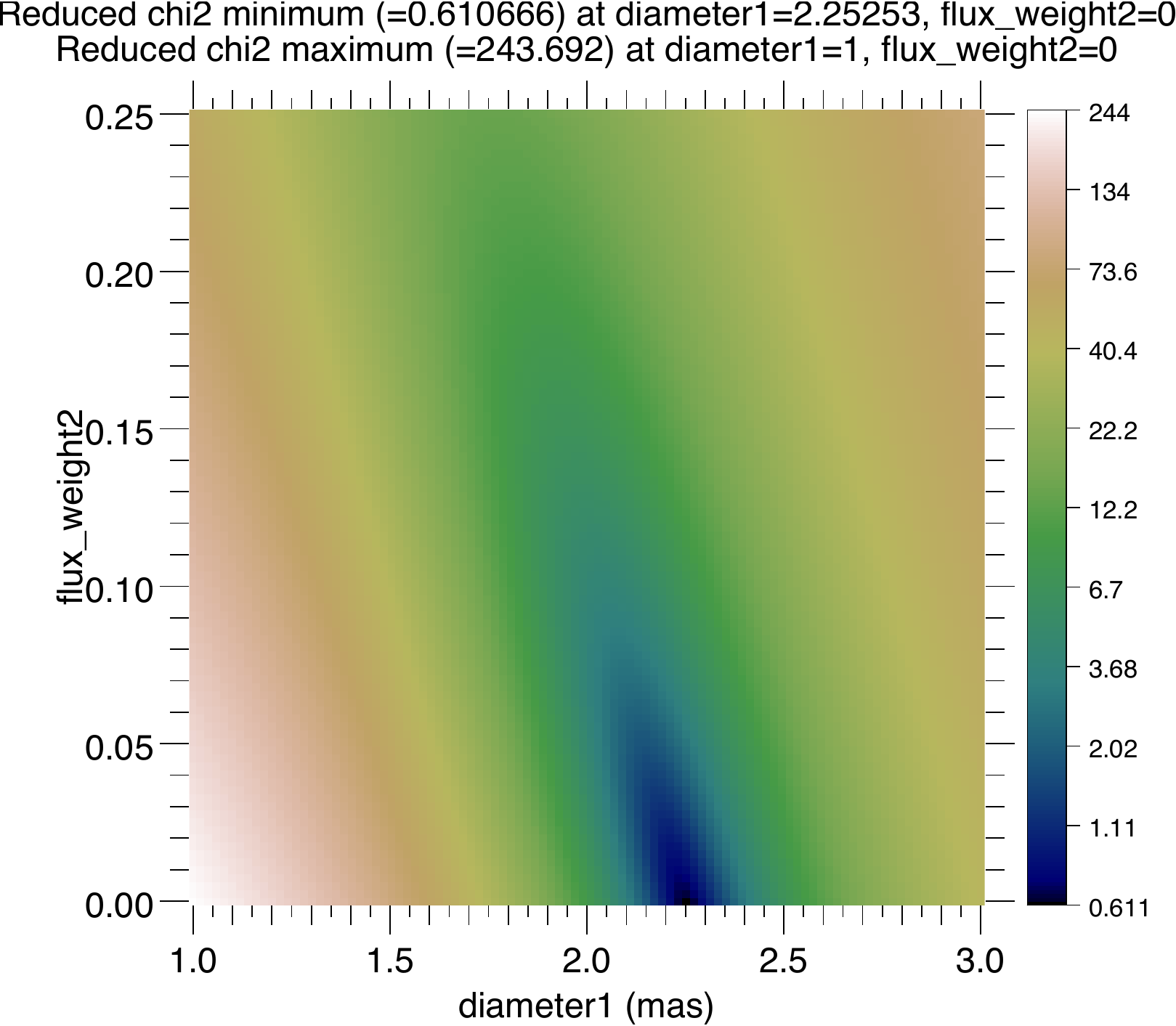}
\caption{\label{Fig:chi2mapV1261}
Same as Fig.~\ref{Fig:chi2map190} for V1261~Ori.}
\end{figure}

\begin{figure}
\includegraphics[width=9cm]{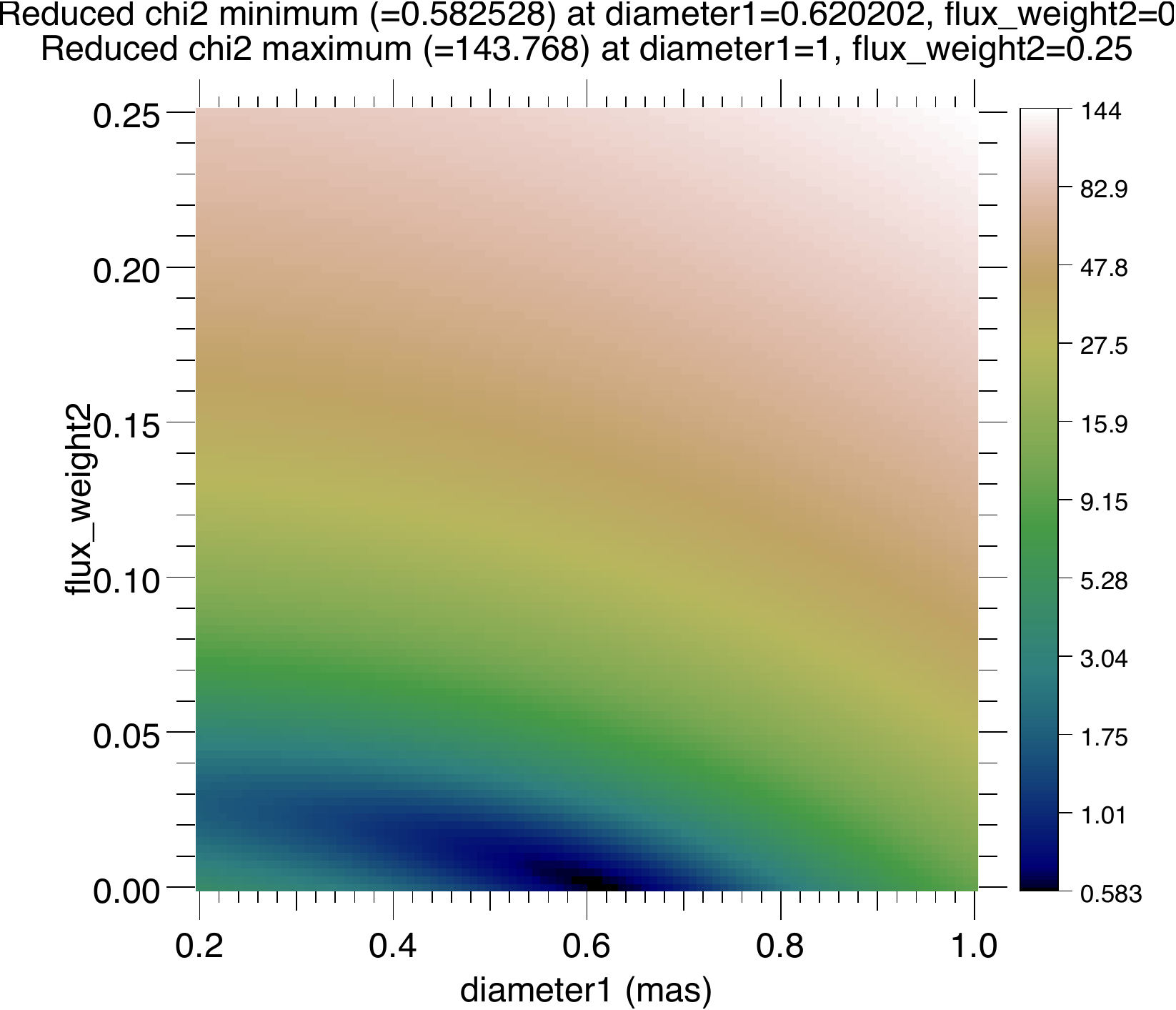}
\caption{\label{Fig:chi2maperdel}
Same as Fig.~\ref{Fig:chi2map190} for ER~Del.}
\end{figure}

\begin{figure}
\includegraphics[width=9cm]{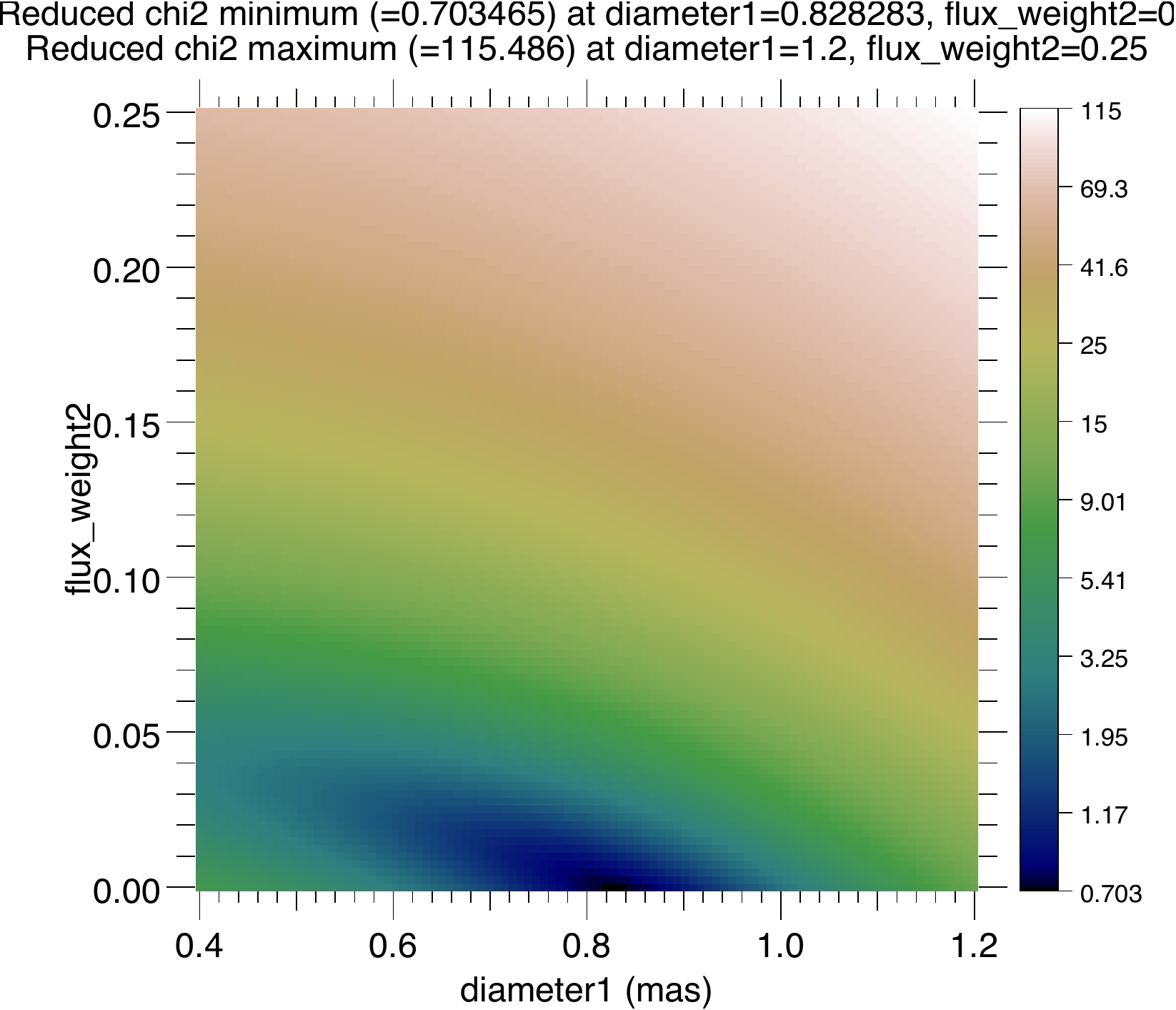}
\caption{\label{Fig:chi2mapfgserj}
Same as Fig.~\ref{Fig:chi2map190} for FG~Ser as observed on 3 July 2012.}
\end{figure}

\begin{figure}
\includegraphics[width=9cm]{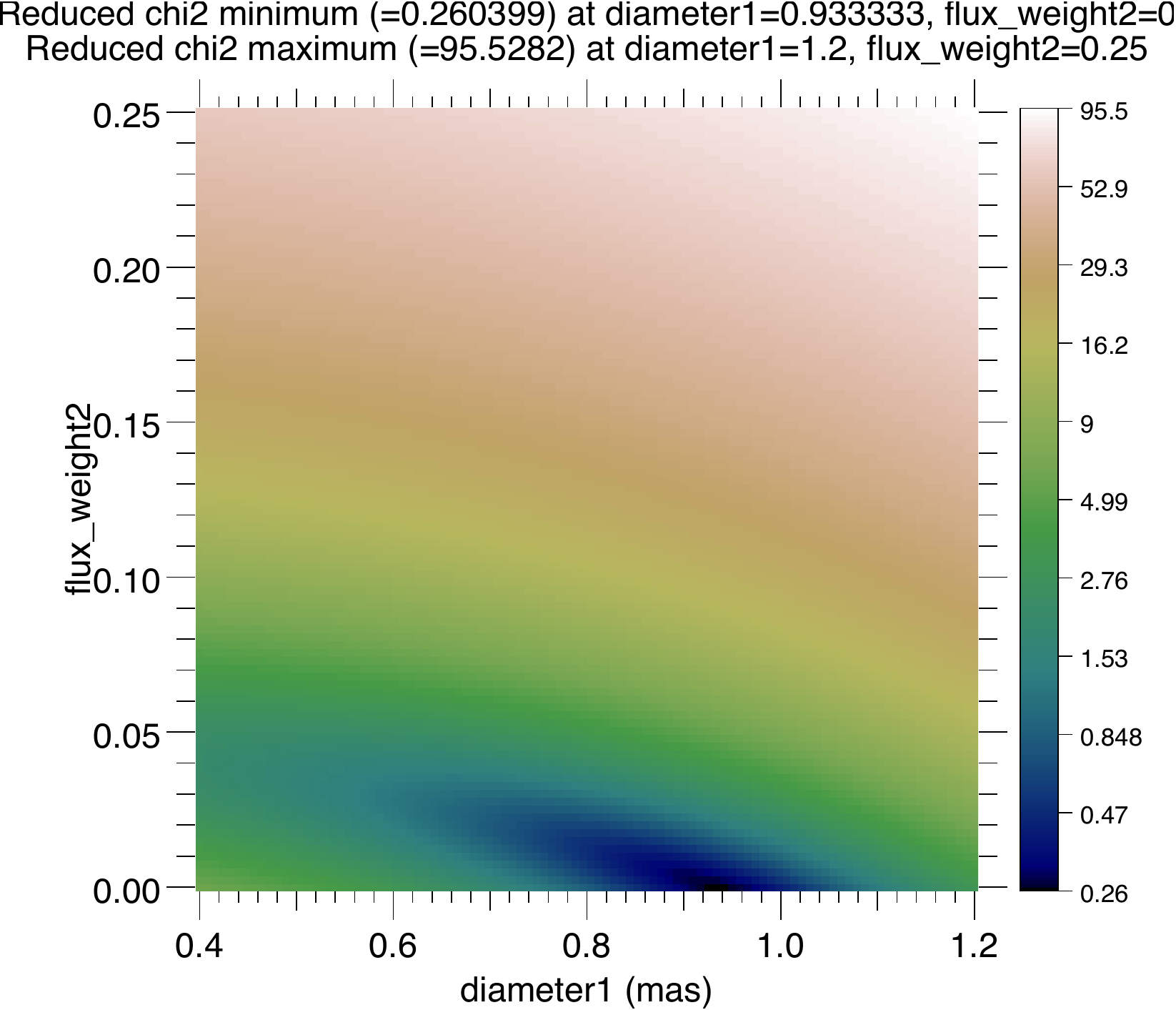}
\caption{\label{Fig:chi2mapfgsera}
Same as Fig.~\ref{Fig:chi2map190} for FG~Ser as observed on 13 August 2012.}
\end{figure}

\begin{figure}
\includegraphics[width=9cm]{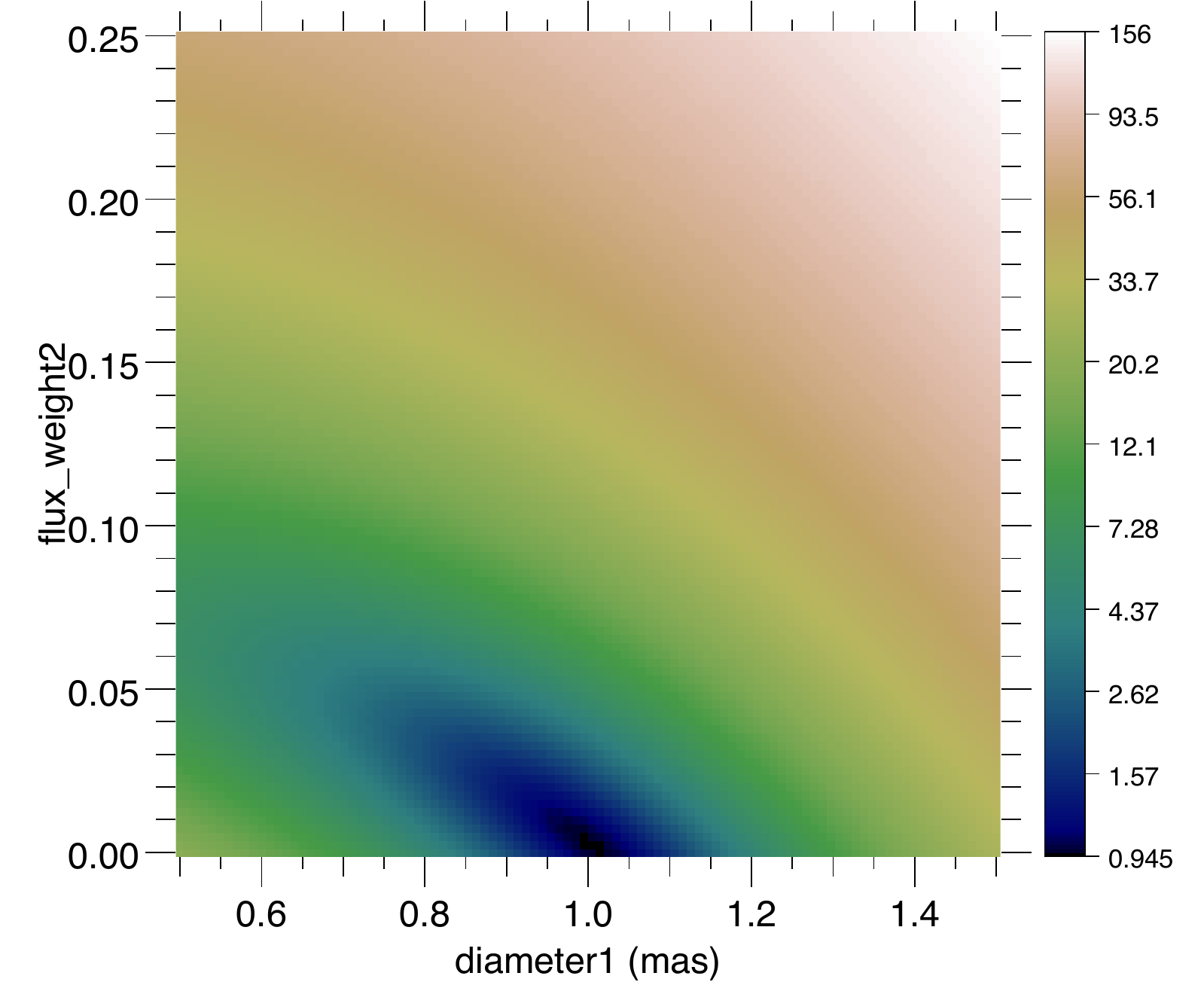}
\caption{\label{Fig:chi2mapagpeg}
Same as Fig.~\ref{Fig:chi2map190} for AG~Peg.}
\end{figure}

\begin{table*}[htbp]
   \centering
    \caption{  \label{tab:calibPIONIER} Calibrators used for PIONIER observations.}
 
   \begin{tabular}{@{} lrrccccc @{}} 
        \hline\hline
        \multicolumn{8}{c}{Calibrators for FG Ser}\\
        \hline
Calibrator    &   R.A.~~~~~~     &    DEC.~~~~~~      &    UD   & UDerr & SpT. & $V $ & $K$  \\
\hline
HD 168245 &  18 18 59.263& $-$04 06 19.66&0.501&0.036&G7 II& 7.6 & 5.1\\
HD 168744 & 	18 21 22.891& $-$03 06 23.28&0.406&0.029&G8/K0 III&7.6&5.4\\
HD 166583 & 	18 11 12.126&+01 58 54.56&0.579&0.041&K3 III&7.7&4.9\\
\hline
\multicolumn{8}{c}{Calibrator for HD~190658}\\
\hline
Calibrator    &   R.A.~~~~~~     &    DEC.~~~~~~      &    UD   & UDerr & SpT. & $V $ & $K$  \\
\hline
HD 185758 & 19 40 05.791&+18 00 50.01&1.352&0.096&G0 III&4.4&2.7\\
\hline
\multicolumn{8}{c}{Calibrators for AG Peg}\\
\hline
Calibrator    &   R.A.~~~~~~     &    DEC.~~~~~~      &    UD   & UDerr & SpT. & $V $ & $K$  \\
\hline
HD 207980 & 	21 52 50.008&+13 29 19.80&0.539&0.038&K0&7.3&4.9\\
HD 208443 & 	21 55 59.069&+10 05 49.07&0.533&0.038&K0&7.1&4.9\\
HD 209166 & 	22 01 05.350&+13 07 11.36&0.454&0.032&F4 III&5.6&4.8\\
\hline
\multicolumn{8}{c}{Calibrators for ER Del}\\
\hline
Calibrator    &   R.A.~~~~~~     &    DEC.~~~~~~     &    UD   & UDerr & SpT. & $V $ & $K$  \\
\hline
HD 198166 & 	20 48 21.475&+07 35 39.21&0.517&0.037&K0&7.1&4.9\\
HD 199255 & 	20 55 47.054&+07 39 58.506&0.422&0.030&G5&7.4&5.3\\
\hline
\multicolumn{8}{c}{Calibrators for HD~352}\\
\hline
Calibrator    &   R.A.~~~~~~     &    DEC.~~~~~~      &    UD   & UDerr & SpT. & $V $ & $K$  \\
\hline
HD 6 &		00 05 03.823&$-$00 30 10.93&0.793&0.056&G9 III&6.3&3.8\\
HD 587&		00 10 18.870&$-$05 14 54.92&1.022&0.073&K1 III&5.8&3.4\\
\hline
\multicolumn{8}{c}{Calibrator for V1261 Ori}\\
\hline
Calibrator    &   R.A.~~~~~~     &    DEC.~~~~~~      &    UD   & UDerr & SpT. & $V $ & $K$  \\
\hline
HD 36134 & 05 29 23.686 & $-$03 26 47.02 & 1.166 & 0.083 & K1 III& 5.8 & 3.2\\
\hline
 \end{tabular}

UD=uniform disc diameter in mas; UDerr=error on UD; SpT=Spectral Type; $V$ and $K$ are the magnitudes in these bands.
 \end{table*}

\end{document}